\documentclass[twocolumn,showpacs,superscriptaddress,preprintnumbers,amsmath,amssymb]{revtex4-2}
\usepackage{amssymb}		
\usepackage{amsmath}
\usepackage{graphicx}
\usepackage[normalem]{ulem}
\usepackage{multirow}
\usepackage{appendix}
\usepackage{CJK}
\usepackage[usenames]{color}
\usepackage{bm}
\usepackage[colorlinks,linkcolor=blue,urlcolor=blue,citecolor=blue]{hyperref}
\usepackage{booktabs} 	
\usepackage{leftidx}

\usepackage{ulem}
\newcommand\redsout{\bgroup\markoverwith{\textcolor{red}{\rule[0.5ex]{2pt}{0.4pt}}}\ULon}

\usepackage[all]{hypcap} 
\makeatletter

\newcommand{\Rmnum}[1]{\expandafter\@slowromancap\romannumeral #1@}
\makeatother

\usepackage{tikz,xcolor,hyperref}

\definecolor{lime}{HTML}{A6CE39}
\DeclareRobustCommand{\orcidicon}{
	\begin{tikzpicture}
	\draw[lime, fill=lime] (0,0) 
	circle [radius=0.16] 
	node[white] {{\fontfamily{qag}\selectfont \tiny ID}};
	\draw[white, fill=white] (-0.0625,0.095) 
	circle [radius=0.007];
	\end{tikzpicture}
	\hspace{-2mm}
}
\foreach \x in {A, ..., Z}{%
	\expandafter\xdef\csname orcid\x\endcsname{\noexpand\href{https://orcid.org/\csname orcidauthor\x\endcsname}{\noexpand\orcidicon}}
}

\begin{document}
	\begin{CJK*} {UTF8} {gbsn}

\title{Effects of the $\alpha$-cluster  structure and the intrinsic momentum component of nuclei on the longitudinal asymmetry in relativistic heavy-ion collisions}

\author{Ru-Xin Cao(曹汝鑫)}

\affiliation{Shanghai Institute of Applied Physics, Chinese Academy of Sciences, Shanghai 201800, China}
\affiliation{Key Laboratory of Nuclear Physics and Ion-beam Application (MOE), Institute of Modern Physics, Fudan University, Shanghai 200433, China}
\affiliation{School of Nuclear Sciences and Technology, University of Chinese Academy of Sciences, Beijing 100049, China}

\author{Song Zhang(张松)\orcidB{}
}
\thanks{Email: song\_zhang@fudan.edu.cn}
\affiliation{Key Laboratory of Nuclear Physics and Ion-beam Application (MOE), Institute of Modern Physics, Fudan University, Shanghai 200433, China}
\affiliation{Shanghai Research Center for Theoretical Nuclear Physics， NSFC and Fudan University, Shanghai 200438, China}

\author{Yu-Gang Ma(马余刚)\orcidC{}}\thanks{Email:  mayugang@fudan.edu.cn}
\affiliation{Key Laboratory of Nuclear Physics and Ion-beam Application (MOE), Institute of Modern Physics, Fudan University, Shanghai 200433, China}
\affiliation{Shanghai Research Center for Theoretical Nuclear Physics， NSFC and Fudan University, Shanghai 200438, China}
\date{\today}

\begin{abstract}
The longitudinal asymmetry in relativistic heavy ion collisions arises from the fluctuation in the number of nucleons involved. This asymmetry causes a rapidity shift in the centre of mass of the participating zone. Both the rapidity shift and the longitudinal asymmetry have been found to be significant at the top LHC energy for collisions of identical nuclei, and the longitudinal asymmetry is important for reconstructing the colliding vertex and correcting the rapidity shift. However, much discussion of the longitudinal asymmetry has treated the initial condition as a non-zero momentum contributed only by the number of participants, i.e. the asymmetry depends only on the number of participating nucleons. { So we naturally raise a physical problem, can other initial conditions, such as two typical initial conditions for nuclei, geometric configuration and momentum distribution, provide effects on the longitudinal asymmetry?
Therefore, in this work we consider other effects on the longitudinal asymmetry other than the fluctuation in the number of participants}, e.g. the $\alpha$ clustering structure as well as the intrinsic momentum distribution in the target and projectile nuclei for the collisions in the framework of a multiphase transport (AMPT) model. By introducing systems with different $\alpha$-clustering structure and intrinsic momentum distribution, we calculated the ratio of the rapidity distributions of different systems and extracted expansion coefficients to analyse the difference contributed by these factors. And we investigated the possible effect of the non-Gaussian distribution on the rapidity distribution. {These results can help us to constrain the initial conditions and reconstruct the colliding vertex in ultra-relativistic heavy ion collisions, and suggest a quantitative correction on the final state measurement and a possible correlation between the initial condition and the final state observable in LHC and RHIC energy.}
\end{abstract}
\maketitle

\section{Introduction}
For decades, relativistic heavy-ion collision experiment has been an important approach to study properties of strong interaction as well as quark-gluon plasma (QGP) which was supposed existed in the  early universe 
\cite{RMP,PhysRep1,PhysRep2,PhysRep3,PRL1,PRL2,ZhuLL,ZhuLL2,SSS,Huang,Luo,Ko,Sun,Rapp}. 
Generally in relativistic heavy-ion collisions, we treat colliding nucleons as two parts, i.e. participants that take part in collisions and spectators that simply pass through the collision zone without interaction. For a collision between non-identical nuclei, the number of participating nucleons from each nucleus is naturally different. However, for a collision between identical nuclei, the number of participants may also fluctuate event-by-event. That means the numbers of participants in two colliding nuclei may also lead to an inequality. This inequality from participant number fluctuation will lead to a non-zero net momentum of the nucleon-nucleon centre of mass frame, but assumed fixed momentum for each nucleon in advance. Thus the center of mass of participants will shift from the collider center of mass of the system, further results in the rapidity shift at final state. This  effect  was usually called  as longitudinal asymmetry \cite{longi_Thakur,PhysRevC.97.024912}. The longitudinal asymmetry reflects the fluctuation of nucleon at initial state, and may manifest in some phenomena. For instance, the $\Lambda$ polarization was investigated in Ref. \cite{Deng:2021miw} which applied the Ultrarelativistic Quantum Molecular Dynamics (UrQMD) model \cite{UrQMD,UrQMD1,UrQMD2,Xi} and gave global spin polarization of $\Lambda$ hyperon for $^{108}$Ag + $^{108}$Ag and $^{197}$Au + $^{197}$Au collisions at $\sqrt{s_{NN}} = $2.42-62.4 GeV. In that work it was compared with measurements from the HADES Collaboration \cite{HADES} and STAR Collaboration \cite{PhysRevC.104.L061901} and fitted well at lower energies. They concluded that the global polarization
was a result of the global angular momentum of the system, so that the longitudinal asymmetry involving initial momentum spatial asymmetry may also be correlated to the polarization phenomena.

\par Previous study on longitudinal asymmetry usually focus on the effects from participant fluctuation between target and projectile. {Thus a motivation naturally arises, can other effects at initial state of collision provide additional significant contribution to longitudinal asymmetry? 
Based on this motivation, we consider two important effects at initial state -- $\alpha$-clustering structure in light nuclei and short range correlation, which may intensify the longitudinal asymmetry.}

\par $\alpha$-clustered nucleus was proposed by Gamow \cite{article_Gamow}, which can be regarded as a special case of nuclear structure.
In that view, in stable nuclei especially for 4$N$ nuclei, some small groups (like $\alpha$) made up of two protons and two neutrons are likely to exist. Then in the nucleus these groups are connected in different shapes like triangle in $^{12}\mathrm{C}$, tetrahedron in $^{16}\mathrm{O}$ and so on. The clustering effect is important to nuclear equation of state, nucleosynthesis and many other problems \cite{Qin,He:2014iqa,Ma_ML,Ma_NST}.  Various observables have therefore been proposed to study the clustering of nuclei in the heavy-ion reaction, such as collective flow \cite{PhysRevLett.112.112501,Li:2020vrg,Guo:2017tco}, multiplicity correlation \cite{Li:2022bpm,Li:2021znq}.  A recent review can be found in \cite{Ma:2022dbh,Ma_NuclTech}. So we assume that such geometry configurations are likely to affect the fluctuation of numbers of participants at initial state, and further contribute to the longitudinal asymmetry. 

\par Another effect taken into our account is short range correlation (SRC). The SRC can partly arise from the nucleon-nucleon short-range central interaction \cite{Antonov1980,CiofidegliAtti:1991mm}. And the intrinsic momentum distribution of nucleons is a direct reflection, which shows us the probability to find a nucleon at certain momentum in a nucleus. When using high values of nucleon momentum and removal energy to describe nucleon spectral function, the function can be written in the form of a convolution integral involving the momentum distributions describing the relative and center-of-mass motion of a correlated nucleon-nucleon pair embedded in the medium \cite{CiofidegliAtti:1991mm}. High momentum tail (HMT), as a direct result from SRC, can be found in momentum distribution of nucleons, and some studies show that the contribution of HMT is mainly provided by proton-neutron pairs \cite{CiofidegliAtti:1991mm,Wang:2017odj}. In Ref.~\cite{Shen:2021dll} the related phenomenon in an Extended Quantum Molecular Dynamics (EQMD) model has been discussed, and the effects on emission time distribution, momentum spectrum and momentum correlation function of two emitted protons of $^{12}\mathrm{C}$-$^{11}\mathrm{B}$ reaction are also investigated, which demonstrated the importance of SRC. The intrinsic momentum distribution of nucleon may also affect shift of initial center of mass, then affect the longitudinal asymmetry.

\par Under AMPT frame, it is simulated that $^{12}$C + $^{12}$C collisions with/without $\alpha$-cluster at center of mass energy $\sqrt{s_{NN}}=$ $6.37$ TeV and $200$ GeV, $^{12}$C + $^{12}$C collisions with/without intrinsic momentum distribution at 200 GeV, as well as $^{197}$Au + $^{197}$Au collisions with Woods-Saxon configuration and high-momentum-tail configuration at 200 GeV. The $0\sim10\%$ centrality is always adopted in all simulations. With the same $\sqrt{s_{NN}}$ and configuration (such as the default Woods-Saxon), comparison between different systems, for example, C  + C and Au + Au, reveals the system size dependence of longitudinal asymmetry. Also for the same configuration like Woods-Saxon, comparison between at 200 GeV and 6.37TeV in C + C collisions shows us the energy dependence of longitudinal asymmetry. Similarly, at the same $\sqrt{s_{NN}}$, comparison between systems with Woods-Saxon and $\alpha$-cluster reveals effect on longitudinal asymmetry from geometry configuration, comparison between systems with Free-Fermi-Gas and High-Momentum-Tail reveals effect on longitudinal asymmetry from intrinsic momentum distribution, in which the High-Momentum-Tail case can show us how the short range correlation in nucleon pair change longitudinal asymmetry.

The paper is organised as follows: in Sec.~\ref{sec:Models_and_Methods}, we gave brief introductions of the models used in our simulation - the AMPT model, the $\alpha$ cluster structure and the HMT effect. Then we introduced basic methods to calculate these longitudinal asymmetry parameters and to provide correction of our $\alpha$-cluster effect and HMT effect. We also suggested possible reasons to explain the differences between different results, and linked these reasons to some further investigations in later works. In Sec.~\ref{sec:results}, we used AMPT to simulate C + C and Au + Au collisions with different initial conditions, and extracted their longitudinal asymmetry parameters and expansion coefficients. We then compared the parameters and coefficients from different systems and pointed out their differences. In Sec.~\ref{sec:explanation}, we explained the effect on longitudinal asymmetry from the initial condition, which can give us insights and guidance on how to constrain the collision conditions, reconstruct the colliding vertex, and relate the observed final state to the effect of different systems in future experimental measurements. Finally, in Sec.~\ref{sec:summary} we give the conclusion and outlook of our work. 

\section{Models and Methods}
\label{sec:Models_and_Methods}
\subsection{Introduction to AMPT}

A multiphase transport model~\cite{AMPT2005,AMPTGLM2016,AMPT2021} is composed of four stages to simulate relativistic heavy-ion collisions. It has successfully described various phenomena at RHIC and LHC energies and becomes a well-known event generator. The AMPT has two versions: String Melting (SM) and Default. In SM version, Heavy Ion Jet Interaction Generator (HIJING)~\cite{HIJING-1,HIJING-2} is used to simulate the initial conditions, then Zhang's Parton Cascade (ZPC)~\cite{ZPCModel} is used to describe interactions for partons which are from all of hadrons in the HIJING but spectators, after which a simple Quark Coalescence Model describes hadronization process, finally A Relativistic Transport (ART) model~\cite{ARTModel} simulates hadron re-scattering process. The Default version of AMPT only conducts the minijet partons in partonic scatterings via ZPC and uses the Lund string fragmentation to perform hadronization.

 AMPT model  ~\cite{AMPT2005,AMPT2021} can describe the $p_T$ spectrum and energy dependence of identified particles such as pion, kaon, $\phi$, proton and $\Omega$ produced in heavy-ion collisions ~\cite{AMPTGLM2016,Jin2018,Wang2019},  as well as the collective flows and temperature during evolution etc~\cite{WangH0, AMPT_temperature_parton_Lin,NSTSongFlow,PhysRevC_Cao_2022,Chen_NT}.  Chiral and magnetic related anomalous phenomena can also be described by the AMPT model \cite{ZhaoXL2019,Gao2020,Huang2020,WangCZ2021,Zhao2019,WuWY}. Further details of the model description and the selection of the parameter set can be found in Refs.~\cite{AMPT2005,AMPTGLM2016,AMPT2021}. 

\subsection{$\alpha$-cluster structure}

In recent several decades, various theoretical models were developed to study the $\alpha$-cluster structure, such as the Fermion Molecular Dynamics model (FMD) \cite{Feldmeier:1989st,Chernykh:2007zz}, the Antisymmetric Molecular Dynamics model (AMD) \cite{Kanada-Enyo:2012yif,Kanada-Enyo:2005
}, the extended Quantum Molecular Dynamics model (EQMD) \cite{He:2014iqa,He:2016cwt,Huang:2017ysr,Wangshanshan:2015} and so on. In our simulation, the initial nucleon distribution in nuclei is configured in the HIJING model with either a pattern of Woods-Saxon distribution or an exotic nucleon distribution which is embedded to study the $\alpha$-clustered structure \cite{He:2014iqa,Li:2020vrg}. The parameters set for the triangle structure are inherited from an extended quantum molecular dynamics (EQMD) model~\cite{He:2014iqa}. EQMD is extended from the quantum molecular dynamics (QMD) model, which can give reasonable $\alpha$-cluster configurations for 4N nuclei  by taking the effective Pauli potential and dynamical wave packet into account. And more details for parameter setting can be seen in Ref.~\cite{He:2014iqa,Li:2020vrg}.

\subsection{High momentum component (HMT)}

The high-momentum-tail  caused by short range correlation  is also proposed to contribute to the longitudinal asymmetry in heavy-ion collisions. By comparing calculated results from model with inclusive and exclusive experiments \cite{CiofidegliAtti:1991mm,Wang:2017odj,Patsyuk:2021fju,CiofidegliAtti:1995qe}, the momentum distribution can be described as two parts: $n_0(k)$ corresponding to low-momentum part which is dominated by single particle features of nucleon structure, $n_1(k)$ corresponding to high-momentum part which is dominated by short-range properties of nucleon structure. In a simple way, one can write the momentum distribution as~\cite{CiofidegliAtti:1995qe}:
\begin{eqnarray}
\left\lbrace
\begin{aligned}
 &n(k)\approxeq n_{0}(k)=\frac{1}{4\pi A}\sum_{\alpha<\alpha_{F}}{A_{\alpha}n_{\alpha}(k)} \text{ for $k<\hat{k}$}\\
 &n(k)\approxeq n_{1}(k)= C^{A}n_{deut}(k) \text{ for $k>\hat{k}$}
\end{aligned},
\right.
\label{eq:momentum_distribution}
\end{eqnarray}
where the subscript $F$ in $\alpha_{F}$  means Fermi level and Fermi momentum, and other variables can all be parameterized from light nuclei momentum distribution fitting \cite{CiofidegliAtti:1995qe}.  For the above distribution, it is always compared with Free-Fermi-Gas (FFG) distribution in this work. More details for parameterization can be found in Ref.~\cite{CiofidegliAtti:1995qe}. In this work, we add this distribution into initialization of AMPT model. 
The default case is the Woods-Saxon distribution, which generally describes only the potential of the nucleon. The FFG case means free Fermi gas, where the momentum distribution of all nucleons is below the Fermi momentum. However, for our focus -- HMT, the nucleon's momentum could reach a high momentum tail, corresponding to $n_{1}(k)$ resulting from SRC. 

\subsection{Methodology}

 Generally, the longitudinal asymmetry can be characterized by some parameters \cite{longi_Thakur}. Here we give the rapidity shift $y_{0}$, asymmetry of participants $\alpha_{part}$ and asymmetry of spectator $\alpha_{spec}$:
\begin{eqnarray}
 &y_{0}=\frac{1}{2}ln\frac{A}{B}\label{eq:y0},\\
 &\alpha_{part}=\frac{A-B}{A+B}\label{eq:a_part},\\
 &\alpha_{spec}=\frac{(N-A)-(N-B)}{(N-A)+(N-B)}=\frac{B-A}{2N-(A+B)},
\label{eq:a_spec}
\end{eqnarray}
where, $A$ and $B$ mean numbers of nucleon participating from the two colliding nuclei (naturally for identical nuclei $A$ and $B$ are equivalent), and $N$ is the total number of nucleons in each nucleus. And it should be noted that $y_{0}\approxeq \frac{1}{2}ln\frac{A}{B}$ is appropriate when $m_{0}\ll p$, fortunately it is possible in LHC at TeV scale $m_{0}/p<10^{-6}$ and in RHIC at GeV scale $m_{0}/p<10^{-4}$. Hence we can also write the equation as $y_{0}=\frac{1}{2}ln\frac{1+\alpha_{part}}{1-\alpha_{part}}$. Further, when $\alpha_{part}$ is small enough, it is easy to see that $y_{0}\approx \alpha_{part}$.

With these definition we can classify vast events in terms of their $y_{0}$, for each event of nucleus-nucleus collision has its own rapidity shift $y_{0}$ which is only determined by initial $A$ and $B$. And although we can not directly acquire the $A$ and $B$, the practical experiments provide us indirect method: by gaining energy deposited in the zero-degree calorimeters on either side of the interaction vertex in collider experiments \cite{201820}, we can measure the $\alpha_{spec}$, then $y_{0}$ can be calculated through the transformed equation:

\begin{eqnarray}
\begin{aligned}
y_{0}=\frac{1}{2}ln\frac{(A+B)(1+\alpha_{spec})-2N\alpha_{spec}}{(A+B)(1-\alpha_{spec})+2N\alpha_{spec}}.
\end{aligned}
\label{eq:y0_a_spec}
\end{eqnarray}

\par And further, to keep consistent to the measurement $\alpha_{ZN}$ in ALICE experiments \cite{201820}, the longitudinal asymmetry can also be defined by number of neutrons in spectators, denoted as $A_{spec}^{n}$ and $B_{spec}^{n}$, instead of $\alpha_{spec}$:
\begin{eqnarray}
 &\alpha_{ZN} = \frac{A_{spec}^{n}-B_{spec}^{n}}{A_{spec}^{n}+B_{spec}^{n}} .
\label{eq:alpha_ZN}
\end{eqnarray}

\par
\begin{figure*}[htbp]
	\centering
	\includegraphics[angle=0,width=18.5cm,height=9cm]{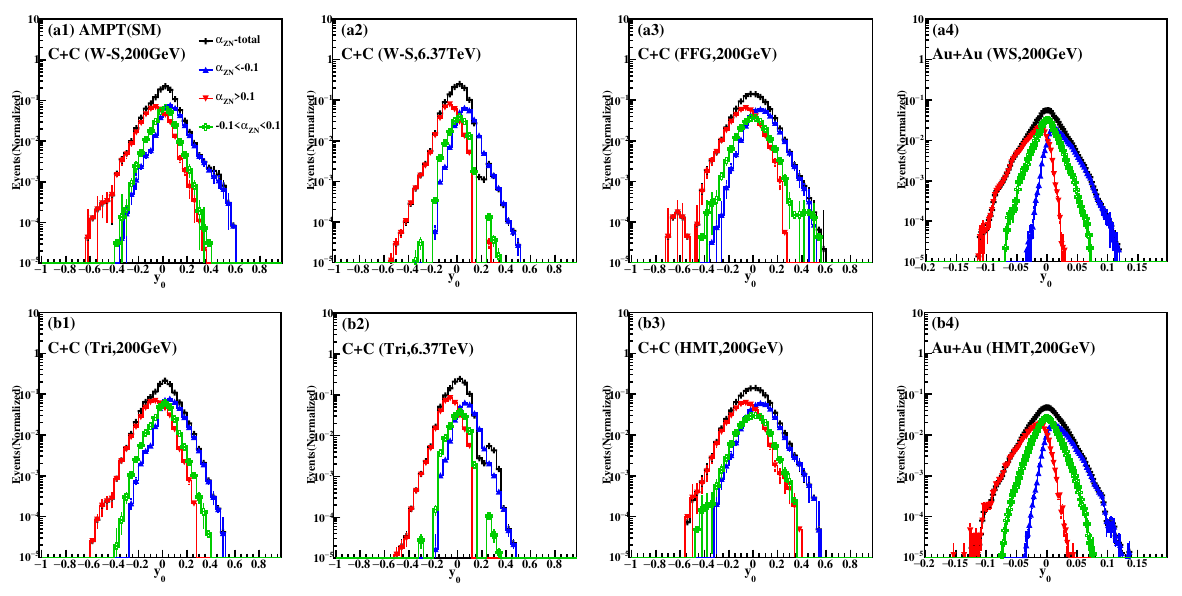}
	\caption{Distribution of parameter $y_{0}$ in different $\alpha_{ZN}$ region for C + C at 200 GeV, C + C at 6.37TeV and Au + Au at 200 GeV with $0\sim10\%$ centrality in AMPT (String Melting) frame.}  
	\label{fig:Fig1_y0_alpha_ZN}
\end{figure*}

 In Fig.~\ref{fig:Fig1_y0_alpha_ZN}, according to different $\alpha_{ZN}$ region \cite{201820}, we plot $y_{0}$ distribution in Au + Au (Woods-Saxon case {and HMT cases}), C + C (Woods-Saxon, FFG and HMT case) collisions at center of mass energy $\sqrt{s_{NN}}$ = 200 GeV and C + C (Woods-Saxon and Triangle case) at $\sqrt{s_{NN}}$ = 6.37 TeV by using AMPT (String Melting) model {and the distribution is consistent with other models' simulation at RHIC or LHC energy }\cite{longi_Thakur,PhysRevC.97.024912,201820}. 

In the distribution of $y_{0}$ shown in Fig.~\ref{fig:Fig1_y0_alpha_ZN}, we should note that if the nucleon intrinsic momentum distribution in the nuclei is taken into account, the definition of rapidity shift $y_0$ should be corrected as,
\begin{eqnarray}
\label{eq:y0_mom_correction}
 y_{0} = \frac{1}{2}ln\frac{1+\alpha_{mom}}{1-\alpha_{mom}},
\end{eqnarray}
where $\alpha_{mom}=\frac{\lvert P_{z}^A\rvert-\lvert P_{z}^B\rvert}{\lvert P_{z}^A\rvert+\lvert P_{z}^B\rvert}$, $P_z^{A}$ and $P_z^{B}$ are the longitudinal momentum of the participants from the two colliding nuclei. Note that $P_z^{A}$ and $P_z^{B}$ would be not equal to the beam momentum due to the effect of the nucleon intrinsic momentum distribution. Also for FFG and HMT cases, the $\alpha_{ZN}$ which is used to divide positive or negative regions should take momentum distribution into account, the $A^{n}_{spec}$ and $B^{n}_{spec}$ in Eq.~(\ref{eq:alpha_ZN}) should be naturally replaced by $P_{z}^{A_{spec}}$ and $P_{z}^{B_{spec}}$.

Now that we have the $y_{0}$ distribution classified by $\alpha_{ZN}$, the longitudinal asymmetry of the different regions becomes obvious. Naturally, for events in $\alpha_{ZN}\textless-0.1$ region (which we call negative $\alpha_{ZN}$ region), $y_{0}$ distribution shows us a positive shift, also $y_{0}$ distribution for events in $\alpha_{ZN}\textgreater0.1$ region (which we call positive $\alpha_{ZN}$ region) shows us a negative shift, and in $|\alpha_{ZN}|\textless0.1$ region, $y_{0}$ distributed in middle region. This negative correlation between $\alpha_{ZN}$ and $y_{0}$ can be understood from Eq.(\ref{eq:a_spec}), for the behaviour of $y_{0}$ intuitively reveals the physical picture of longitudinal asymmetry. For example, in an event, if $A\textgreater B$, 
we have $y_{0}\textgreater0$ according to Eq.~(\ref{eq:y0}). So the rest neutrons as spectators in projectile (noted as $A^{n}_{spec}$) will generally to be less than the rest neutrons as spectators in target (noted as $B^{n}_{spec}$), thus we have $\alpha_{ZN}\textless0$ according to Eq.~(\ref{eq:alpha_ZN}). 

\par
\begin{figure*}[!htbp]
	\centering
	\includegraphics[angle=0,width=18cm,height=18cm]{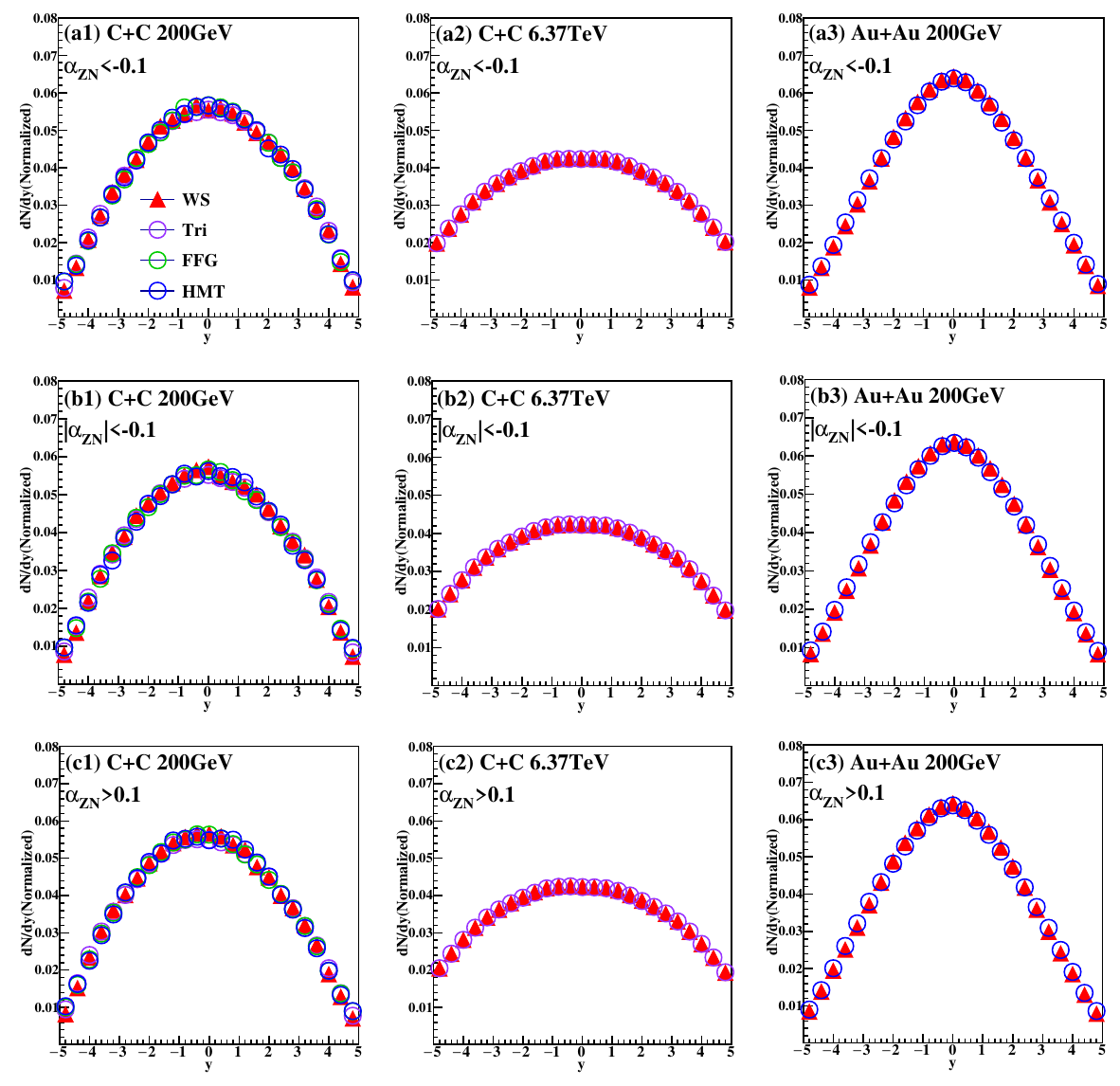}
	\caption{Normalized $dN/dy$ distribution in the positive/middle/negative $y_{0}$ regions for our eight different systems, corresponding to $\alpha_{ZN}<-0.1,-0.1<\alpha_{ZN}<0.1$ and $\alpha_{ZN}>0.1$.}  
	\label{fig:Fig3_rapidity_3_region_pdf}
\end{figure*}

 {Similarity between our $y_{0}$ distribution (especially in Au+Au case) and {the Tuned Glauber Monte Carlo (TGMC) simulation of Pb+Pb case by the ALICE collaboration} can be seen in Ref.~\cite{201820}. And further we can see the charged particles' rapidity distribution of our eight systems in Fig.~\ref{fig:Fig3_rapidity_3_region_pdf}. It seems that our results are close to ideal Gaussian distribution as proposed in Ref.~\cite{longi_Thakur} just by comparing our figures with works in Ref.~\cite{longi_Thakur} at RHIC energy.} {Then more issues beyond ideal cases or experimental results will also be discussed in this work}, by fitting and extraction of $c_{n}$ we will see, besides the ideal Gaussian shape, {the deformation of rapidity shift will also reflect the longitudinal asymmetry}, and fortunately we will disclose that the intrinsic momentum distribution can indeed affect the longitudinal asymmetry by changing the shape of rapidity distribution.

To further investigate the rapidity shift from the longitudinal asymmetry, it is proposed to take the ratio of the rapidity distribution of particles with positive asymmetry to that of negative asymmetry in collisions, $\frac{\left(\frac{dN}{dy}\right)_{+asym}}{\left(\frac{dN}{dy}\right)_{-asym}}$~\cite{PhysRevC.97.024912}, in which the `$+asym$' corresponds to positive $y_{0}$ region $(\alpha_{ZN}<-0.1)$ and `$-asym$' corresponds to negative $y_{0}$ region $(\alpha_{ZN}>0.1)$, so the ratio can be expressed in Taylor expansion,
\begin{eqnarray}
\label{eq:ratio_pos_neg}
\begin{aligned}
\frac{\left(\frac{dN}{dy}\right)_{+asym}}{\left(\frac{dN}{dy}\right)_{-asym}}\propto \sum_{0}^{\infty}c_n y^{n}.
\end{aligned}
\end{eqnarray}
If the rapidity distribution of the particles is in a Gaussian type, $dN/dy \propto \exp\left( -\frac{(y-y_0)^2}{2\sigma^2} \right)$, Eq.~(\ref{eq:ratio_pos_neg}) becomes,
\begin{eqnarray}
\label{eq:ratio_pos_neg_gaussian}
\begin{aligned}
\frac{\left(\frac{dN}{dy}\right)_{+asym}}{\left(\frac{dN}{dy}\right)_{-asym}}\propto \exp{\left(\frac{2yy_{0}}{\sigma^2}\right)}\propto \sum_{0}^{\infty}c_{n}(y_{0},\sigma)y^{n},
\end{aligned}
\end{eqnarray}
where the Taylor expansion coefficients $c_n$ are related to the Gaussian parameters $y_0$ and $\sigma$ and yield $c_{n}(y_{0},\sigma) = \frac{\left(2y_{0}/\sigma^2\right)^n}{n!}$. However， the rapidity distribution of particles does not always follow a Gaussian pattern and the no-Gaussian effect will be discussed later.

\section{Results of longitudinal asymmetry from different systems}

\label{sec:results}
\subsection{$y_{0}$ and numbers of participants}

{Panel (a1)-(a4), (b1)-(b4) in Fig.~\ref{fig:Fig1_y0_alpha_ZN} show the $y_{0}$ distributions at initial sate in C  + C and  Au + Au collisions at $\sqrt{s_{NN}}$ = 200 GeV and C + C collisions at $\sqrt{s_{NN}}$ = 6.37 TeV, respectively, for different $\alpha_{ZN}$ regions. The results are consistent with results for Au + Au simulation at RHIC energy and Pb + Pb measurement in ALICE experiment from various works~\cite{longi_Thakur,PhysRevC.97.024912,201820}.} In this calculation nucleon distributions are configured either as the Woods-Saxon type in $^{12}$C or the $\alpha$-clustered triangle shape in $^{12}$C. The $y_{0}$ distributions in C + C collisions present similar behaviour for the different configurations of the nucleon distribution in the collided nuclei, but show stronger fluctuations than for larger collision systems shown in (a4) and (b4), and also show stronger fluctuations than for larger $\sqrt{s_{NN}}$ in (a2) and (b2). For $y_{0}$ distributions in C + C  collisions with configuration for collided nuclei with nucleon momentum distribution in HMT and FFG. It can be seen the $y_0$ distribution in (a3) and (b3) is affected by the nucleon intrinsic momentum distribution comparing with that in Woods-Saxon distribution in (a1). The former case shows larger width of $y_{0}$ distribution contributed by momentum distribution.

Further in Fig.~\ref{fig:Fig1_y0_alpha_ZN}, by comparing C + C (W-S, 200 GeV) to C + C (W-S, 6.37 TeV), or C + C (Tri., 200 GeV) to C + C (Tri., 6.37 TeV), the systems at higher $\sqrt{s_{NN}}$ (6.37 TeV) show smaller $y_{0}$ fluctuation than those at lower $\sqrt{s_{NN}}$ (200 GeV). {And large system (Au + Au) also shows smaller $y_{0}$ fluctuation than small system (C + C). These physical pictures are consistent with works at RHIC energy in Refs.~\cite{longi_Thakur}. But if we consider initial intrinsic momentum distribution, we can see that $y_{0}$ fluctuation is enhanced by the unfixed momentum in beam direction (in FFG and HMT).} Then in Fig.~\ref{fig:Fig3_rapidity_3_region_pdf}, it can be seen that the rapidity distribution at final state directly corresponds to different $y_{0}$ shift in Fig.~\ref{fig:Fig1_y0_alpha_ZN}. The rapidity distribution with positive shift in $\alpha_{ZN}\textless-0.1$ reflects the positive $y_{0}$ shift in $\alpha_{ZN}\textless-0.1$ and vice versa.

\subsection{Expansion coefficient}

\begin{figure*}[!htbp]
	\centering
	\includegraphics[angle=0,width=18.6cm,height=6.2cm]{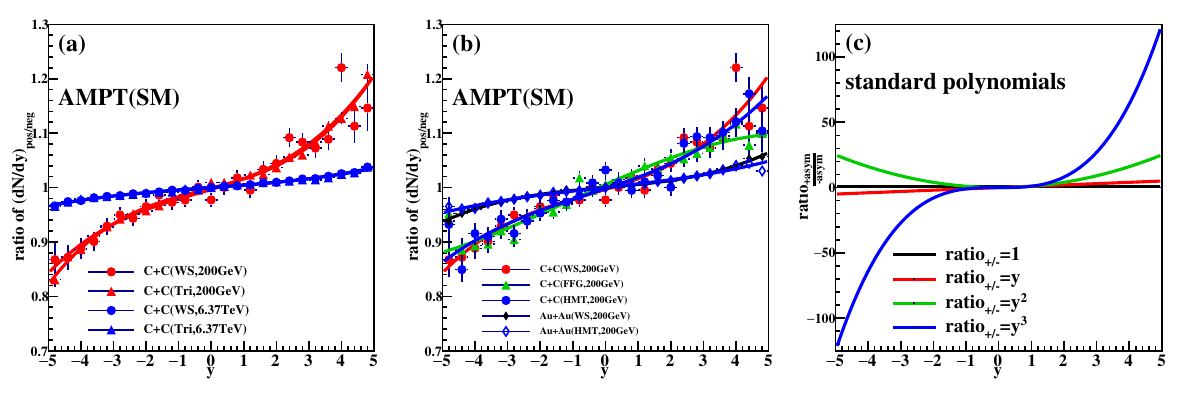}
	\caption{The ratio of $dN/dy$ for our eight different systems, along with fitting curves and standard polynomials for comparison.}  
	\label{fig:Fig4_ratio_fitting}
\end{figure*}

\begin{figure*}[!htbp]
	\centering
	\includegraphics[angle=0,width=18cm,height=18cm]{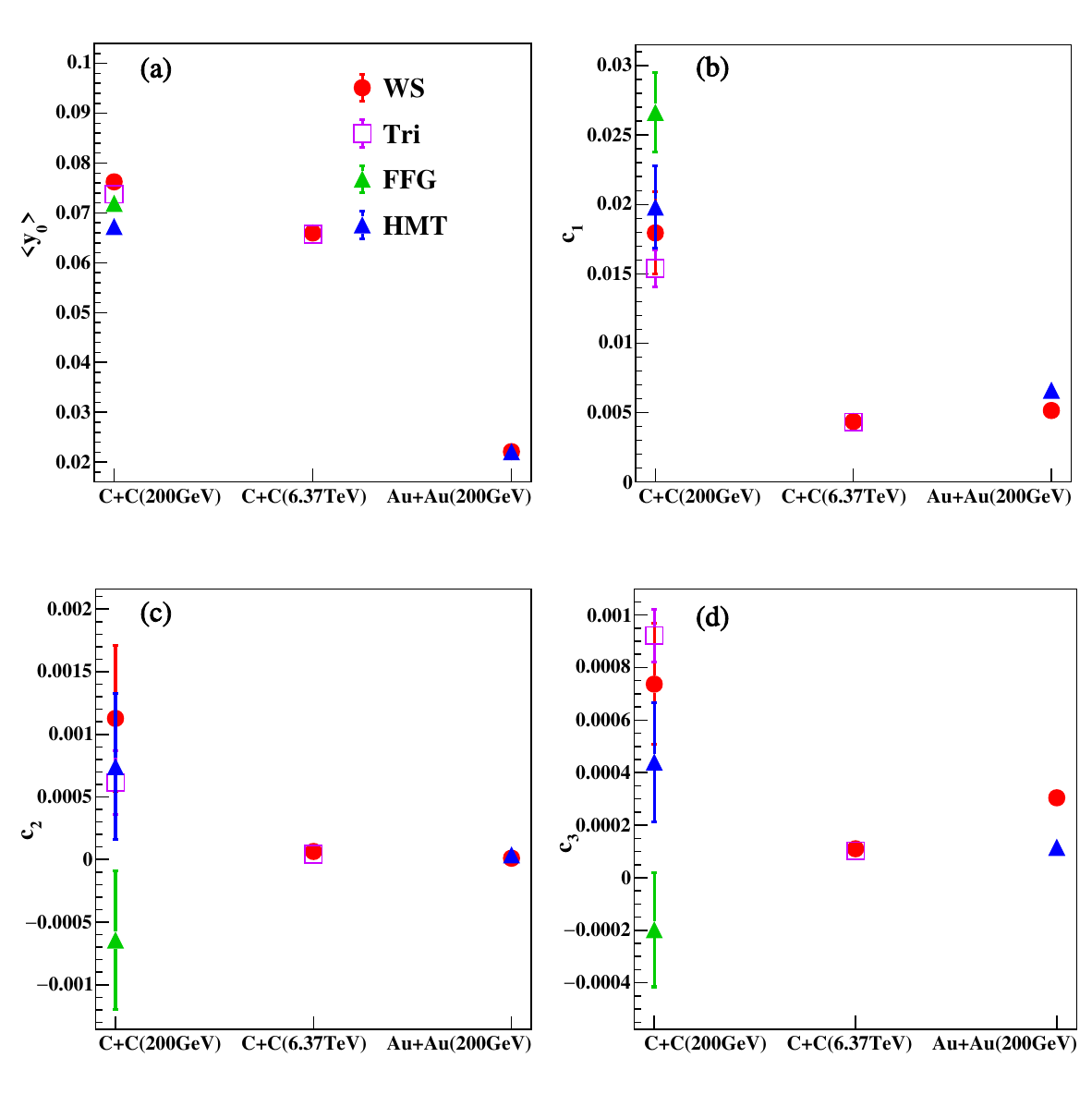}
	\caption{{ $\langle y_{0}\rangle$ and $c_{n}$ extracted from our eight different systems, markers correspond to different configurations.} }
	\label{fig:Fig8_cn_tab}
\end{figure*} 

After plotting initial distribution of parameters, we can calculate $c_{n}$ based on equation~(\ref{eq:ratio_pos_neg}). It is clear that the longitudinal asymmetry becomes harder to be measured as the collision energy increasing or the regions close to middle region ~\cite{longi_Thakur}, thus the later extraction of more parameters like expansion coefficients may become harder to distinguish in investigation. As a result, we choose taking positive and negative region which are far from mid-region so that the events from both sides around symmetry events can provide distinct ratio to investigate further significant properties of longitudinal asymmetry.

The rapidity distributions of charged particles shown in Fig.~\ref{fig:Fig3_rapidity_3_region_pdf} for events from the positive and negative rapidity shift regions in C + C collisions and Au + Au collisions, respectively, for different initial state configurations and collision energies.  To illustrate the longitudinal asymmetry, the differences between the positive and negative shift regions are expressed by the ratio of $\frac{\left(\frac{dN}{dy}\right)_{+asym}}{\left(\frac{dN}{dy}\right)_{-asym}}$ as shown in Fig.~\ref{fig:Fig4_ratio_fitting}. According to Eq.~(\ref{eq:ratio_pos_neg}) and Eq.~(\ref{eq:ratio_pos_neg_gaussian}), a third order polynomial is performed to fit the ratio and the coefficients $c_{0}, c_{1}, c_{2}$ and $c_{3}$ are extracted~\cite{longi_Thakur,PhysRevC.97.024912}. The extracted coefficients $c_{n}$ ($n$ = 0, 1, 2, 3) are plotted in Fig.~\ref{fig:Fig8_cn_tab} for different collision systems with specific initial configurations.

For the $\alpha$ cluster structure case and the Woods-Saxon case in Fig.\ref{fig:Fig8_cn_tab}, at the same $\sqrt{s_{NN}}$, there is no obvious difference between $c_{n}(\mathrm{Tri.})$ and $c_{n}(\mathrm{W-S})$ (here $n$ = 1, 2, 3) within the uncertainty for the same order. If we compare their central values, $c_{1}$ in the triangle case is slightly smaller than $c_{1}$ in the Woods-Saxon case, and $c_{2}$ behaves similarly to $c_{1}$, while $c_{3}$ is larger in the triangle case. In summary, the difference between the Woods-Saxon configuration and the cluster configuration is not clear.

For the case of intrinsic momentum distribution, according to Fig.~\ref{fig:Fig8_cn_tab}, in C + C cases we can see, the first order terms $c_{1}$ in the W-S case are smaller than those in the FFG and HMT cases. However, $c_{2}$ and $c_{3}$ in the W-S case are larger than those in the FFG and HMT cases, respectively, and the high-order terms $c_{2},c_{3}$ in the HMT case are larger than those in the FFG case, even considering their uncertainties. 
{And it is interesting to see that the difference for $c_{n}$ between Au+Au(WS) and Au+Au(HMT) behaves similarly to the C+C system, suggesting to us that HMT may have a similar effect on longitudinal asymmetry in both small and large system sizes.}

\section{Explanation And Further Discussion}
\label{sec:explanation}
\subsection{Ideal Gaussian rapidity distribution and deformed rapidity distribution}

Before discussing the results for $c_{n}$, we should firstly consider how the parameters in ideal Gaussian distribution affect $c_{n}$. According to Eq.~(\ref{eq:ratio_pos_neg_gaussian}), $c_{n}$ can be directly determined by initial shift $y_{0}$ and final rapidity width $\sigma$. 
However, in experiments or transport model simulations, the rapidity distribution does not always have the ideal Gaussian distribution, so that 
Eq.~(\ref{eq:ratio_pos_neg_gaussian}) requires $y_{0}^{+asym}=y_{0}^{-asym}$, $\sigma^{+asym} = \sigma^{-asym}$,  which means that $c_{n}$ is very sensitive to $y_{0},\sigma$ as explained in Ref.~\cite{201820}. We can provide a simple method of estimating the magnitude of the sensitivity. We denote $\frac{\sigma^{+asym}}{\sigma^{-asym}}=m,\frac{y_{0}^{+asym}}{y_{0}^{-asym}}=n$ and choose $\sigma^{+asym} = \sigma,y_{0}^{+asym}=y_{0}$ (just for convenience), the widths and means in Fig.~\ref{fig:Fig3_rapidity_3_region_pdf} give $(m-1)\sim 10^{-3},n\sim 10^{-1}$.  Ignoring small higher order quantities such as $(1-m^{2}),y_{0}^2$, we can estimate the difference between the simulated rapidity distribution and the standard Gaussian shape: $\frac{ratio_{simu}}{ratio_{gaus}}$ $ \sim$ $exp{\frac{m(n+1)yy_{0}}{\sigma^{2}}}$. Both our simulation and Ref.~\cite{201820,longi_Thakur} give $y_{0}\sim 10^{-1},\sigma\in(2,4)$, so we can easily estimate that changing $y_{0}$ and $\sigma$ in the order of $10^{-3}\sim 10^{-1}$ can only lead to $ratio_{simu}$ being about 1.2 times larger than $ratio_{gaus}$. So, besides the sensitivity of $y_{0}$ and $\sigma$, we think that more of the difference of $c_{n}$ is due to the deformation of the rapidity distribution. 

\begin{figure}[htbp]
	\centering
	\includegraphics[angle=0,width=9.5cm,height=6cm]{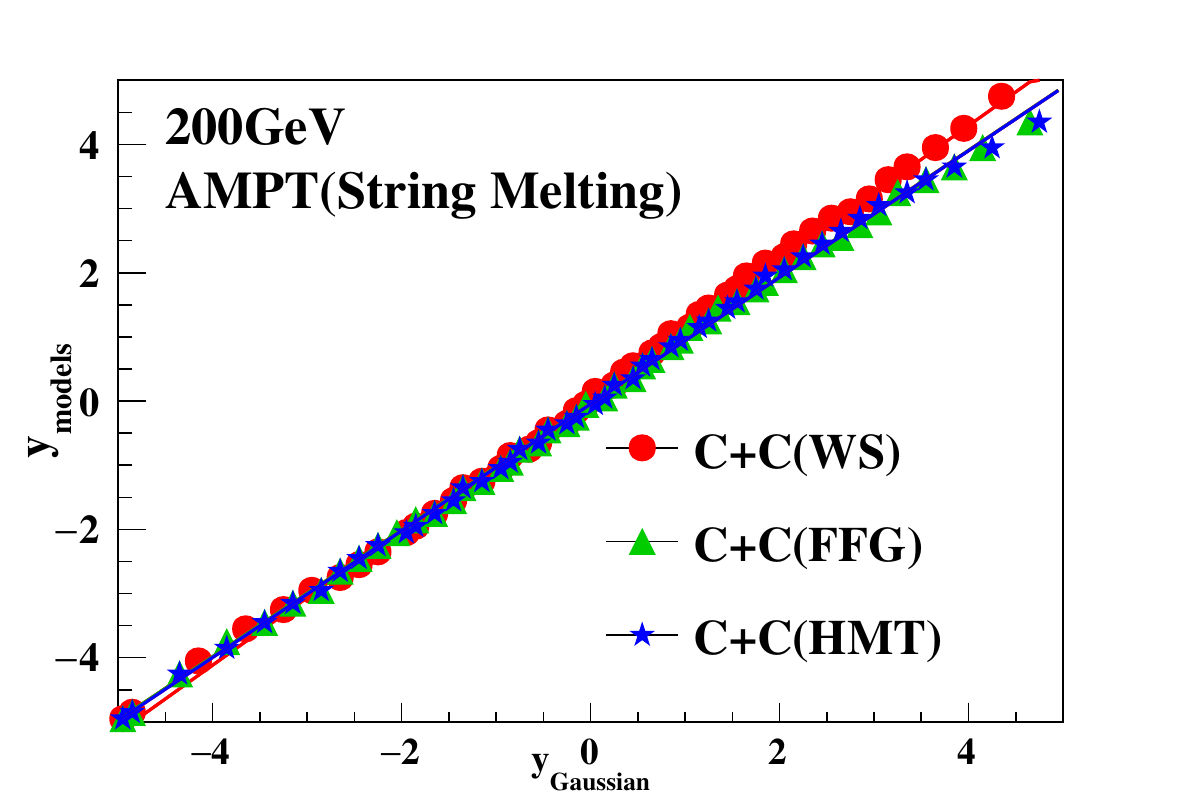}
	\caption{The Q-Q plot to examine normality of systems with different initial momentum and parameterize deformation effect at final rapidity distribution.}  
	\label{fig:Fig7_QQ_plot}
\end{figure} 

Since $c_{n}$ from different initial momentum cases show the most significant difference, we also choose to plot Q-Q plots to compare our W-S, FFG and HMT cases with Gaussian distributions. In statistics, Q-Q plots are usually used to characterise the normality of a given distribution, each distribution has its variable values corresponding to different percentiles, by plotting the scatter of our data sets on the y-axis against the scatter of the Gaussian distribution on the x-axis, we can visually see how close our data sets are distributed to a Gaussian distribution. In general, an approximate linearity like our fitted lines in Fig.~\ref{fig:Fig7_QQ_plot} means that the distribution of our data is close to a Gaussian shape. And meanwhile the intercept shows $y_{0}$ and the slope shows $\sigma$.
~The scatter and fitted lines in Fig.~\ref{fig:Fig7_QQ_plot} do not show any significant difference between the W-S, FFG, HMT cases and the Gaussian distribution. But we can still notice that the rapidity distribution with momentum distribution (FFG and HMT) give different slope and intercept from W-S case, implying to us the effect of intrinsic momentum distribution on rapidity  deformation.

\subsection{Effect on $c_{n}$ from rapidity shift and rapidity deformation in longitudinal asymmetry}

\par
\begin{figure*}[!htbp]
	\centering
	\includegraphics[angle=0,width=18.7cm,height=6cm]{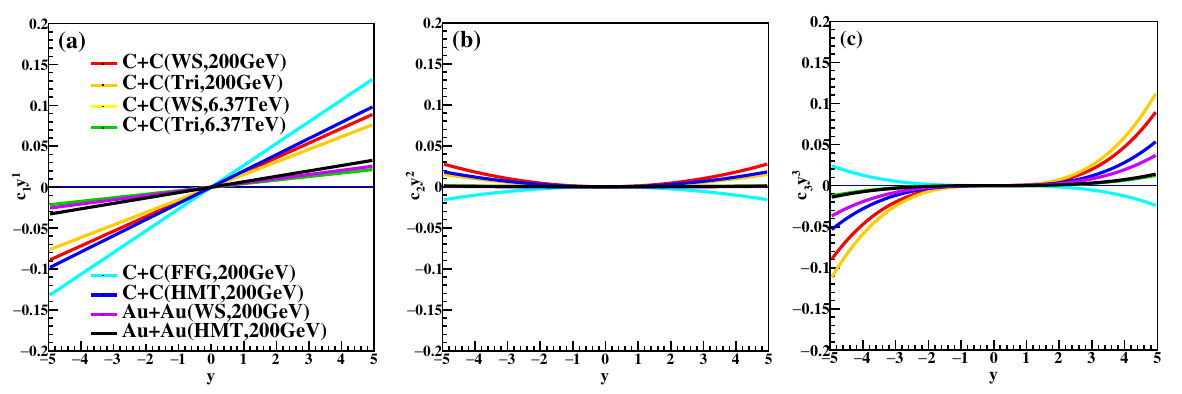}
	\caption{Different components $c_{n}y^{n}$ for our eight systems.}  
	\label{fig:Fig5_cnyn}
\end{figure*}

 Beyond the explanation for the analytic form of the Gaussian distribution, the practical meaning of the expansion coefficient can be understood better from definition of Taylor expansion, that describing function by combination of polynomials. From this point of view, our expansion coefficients $c_{n}$ actually present contribution from powers of rapidity at different orders.  To give a more intuitive explanation, we plot each rapidity ratio along with three standard polynomial: $y,y^{2},y^{3}$ in panel Fig.~\ref{fig:Fig4_ratio_fitting}(c). 
And then we also plot each component $c_{n}y^{n}$ in Fig.~\ref{fig:Fig5_cnyn} to show their contribution to ratio, here different values of $c_{n}$ are shown in Fig.~\ref{fig:Fig8_cn_tab}. 
It is clear that in systems with higher $\sqrt{s_{NN}}$ (C + C, 6.37 TeV) or larger size (Au + Au, 200 GeV), the effect of longitudinal asymmetry is obviously smaller than that in C + C (200 GeV). In Fig.~\ref{fig:Fig5_cnyn} (a), (b) and (c), we can see yellow (C + C, WS, 6.37 TeV), green (C + C, Tri, 6.37 TeV), violet (Au + Au, WS, 200 GeV) {and black (Au+Au, HMT, 200 GeV)} lines are closer to 0 than red (C + C, WS, 200 GeV), orange (C + C, Tri, 200 GeV), cyan (C + C, FFG, 200 GeV), blue (C + C, HMT, 200 GeV) lines, and the longitudinal asymmetry of systems at the same $\sqrt{s_{NN}}$ with different configurations (C + C, WS, 200 GeV in red line and C + C, Tri, 200 GeV in orange line, C + C, WS, 6.37 TeV in yellow line and C + C, Tri, 6.37 TeV in green line) are so close that can hardly be distinguished. So our best choice to discuss how deformation changes the longitudinal asymmetry is to compare C + C (WS, 200 GeV), C + C (FFG, 200 GeV) and C + C (HMT, 200 GeV) systems.

In polynomials we can see, in different region of rapidity, the contribution of $y,y^{2},y^{3}$ are different. As the rapidity $y$ increases from 0 to 1, then to the region greater than 1, the deformation effect contributed by $y^{2}$ and $y^{3}$ becomes more and more significant so that $c_{n}y^{n}$, especially $c_{3}y^{3}$, can be comparable to $c_{1}y^{1}$ as shown in Fig.~\ref{fig:Fig5_cnyn}(a),(c).

\begin{figure*}[!htbp]
	\centering
	\includegraphics[angle=0,width=18.9cm,height=9.5cm]{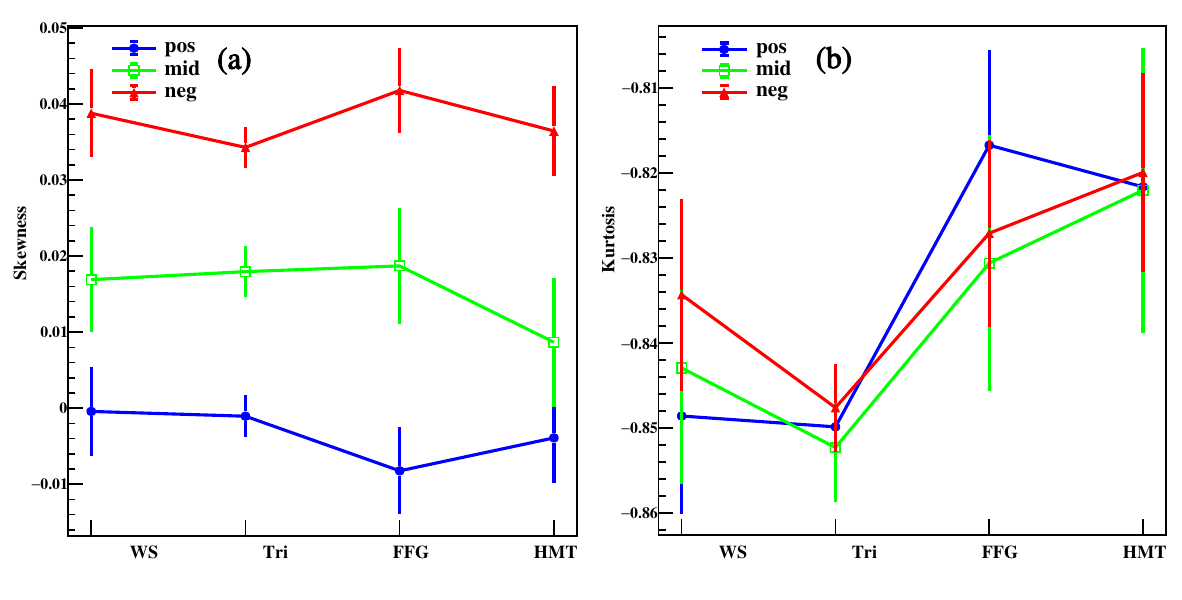}
	\caption{The skewness and kurtosis of rapidity distribution from different configurations and $\alpha_{ZN}$ regions in our eight systems.}  
	\label{fig:Fig6_skewness_kurtosis_new_50bin}
\end{figure*}

 In $-1\textless y\textless1$ region, we have $|y|\textgreater|y^{2}|\textgreater|y^{3}|$, 
 which means the direct rapidity shift $y$ as the linear (also as the leading order) component of the ratio 
 dominates the largest contribution to the $ratio_{+/-}$ in this region. According to Ref.~\cite{201820}, $c_{1}$ shows a linear dependence on $\langle y_{0}\rangle$. {For those cases in which $y_{0}$ only depends on fluctuation of participants (like all the W-S and Tri. cases), $c_{1}$ dependence on $\langle y_{0}\rangle$ is consistent with our expectation and those simulation in ALICE.} For systems at the same $\sqrt{s_{NN}}$ in Woods-Saxon and Triangle case, by comparing $\langle y_{0}\rangle$ with $c_{1}$ in Fig.~\ref{fig:Fig8_cn_tab}, we can see that  $c_{1}$ shows similar linear dependence on $\langle y_{0}\rangle$, and similar dependence can also be shown even in the error (width) of $\langle y_{0}\rangle$ and $c_{1}$ in Fig.~\ref{fig:Fig8_cn_tab}. We can see these $c_{1}$  in $|y|\in(0,1)$ are mainly dominated by rapidity shift.
 
However, when we discuss the region in $y\in(1,5)$, Fig.~\ref{fig:Fig5_cnyn} reminds us that deformation of rapidity distribution also contributes to the ratio, meanwhile for the C+C FFG and HMT cases, their $c_{1}$ dependence on $\langle y_{0}\rangle$ are different from the WS case. In Fig.~\ref{fig:Fig8_cn_tab} we can see FFG (200 GeV) and HMT (200 GeV) have smaller $\langle y_{0}\rangle$ than WS (200 GeV), but larger $c_{1}$ than WS.  In Fig.~\ref{fig:Fig3_rapidity_3_region_pdf}, it is difficult to see the slight deformation intuitively in rapidity distribution. But fortunately, according to Fig.~\ref{fig:Fig1_y0_alpha_ZN}, Fig.~\ref{fig:Fig4_ratio_fitting}, 
Fig.~\ref{fig:Fig5_cnyn} and Fig.~\ref{fig:Fig8_cn_tab}, we can infer how the rapidity distribution deformed at final state in Fig.~\ref{fig:Fig3_rapidity_3_region_pdf}.

For convenience, we can call the region in $|y-\langle y_{0}\rangle|<\langle y_{0}\rangle$ as peak, and the region in $2\langle y_{0}\rangle<|y|<(5-2\langle y_{0}\rangle)$ as ridge. In Fig.~\ref{fig:Fig4_ratio_fitting}(b), we can see, around $y=0$ both C + C (FFG, 200 GeV, green) and C + C (HMT, 200 GeV, blue) show larger ratios than C + C (WS, 200 GeV, red). That means, in $|y|\in(0,1)$ rapidity distribution in FFG and HMT give larger ratios of $\frac{(dN/dy)_{peak}}{(dN/dy)_{ridge}}$ than the WS case (normalized $dN/dy$ has been shown in Fig.~\ref{fig:Fig3_rapidity_3_region_pdf}). This conclusion is a result from deformation of peak and ridge in Fig.~\ref{fig:Fig3_rapidity_3_region_pdf}, and in Fig.~\ref{fig:Fig1_y0_alpha_ZN} we can infer the origin of this deformation. 

In Fig.~\ref{fig:Fig1_y0_alpha_ZN}(a1), (a3) and (b3), we can extract that $y_{0}$ distribution in C +  C (FFG, 200 GeV) and C + C (HMT, 200 GeV) show lower peaks and larger width than that in C + C (WS, 200 GeV), namely, in $\alpha_{ZN}<-0.1$,  $\sigma_{WS} = 0.1011\textless\sigma_{FFG}=0.1016\textless\sigma_{HMT}=0.1057$. These larger widths are caused by additional momentum distribution from FFG and HMT, as we defined in Eq.~\ref{eq:y0_mom_correction}. Hence we see the effect from intrinsic momentum distribution on longitudinal asymmetry at final state. 

But momentum distribution does not only affect $c_{1}$ by causing deformation in $y\in(-1,1)$. In Fig.~\ref{fig:Fig4_ratio_fitting}(b), as y increases to $\pm5$, we can see ratio of C + C (WS, 200 GeV, red) exceeds C + C (FFG, 200 GeV, green) and C + C (HMT, 200 GeV, blue), especially in (-5,-4) and (4,5), after a small peak, the ratios in FFG and HMT case fall closer to 1.00 than WS case. It reminds us that in region close to $\pm5$, rapidity distribution in FFG and HMT are both depressed obviously that the ratios are closer to 1. This depression is a result of deformation at marginal $y$ distribution ($y$$\rightarrow \pm5$), to discuss origin of this deformation, we should go back to check the asymmetry from intrinsic momentum distribution in Fig.~\ref{fig:Fig1_y0_alpha_ZN}. By comparing initial $y_{0}$ distribution in Fig.~\ref{fig:Fig1_y0_alpha_ZN} and final rapidity ratio in Fig.~\ref{fig:Fig4_ratio_fitting}, we can see the asymmetry in both initial and final state is consistent. In $y_{0}$ distribution, FFG and HMT provide larger width around $y_{0}=0$ with less events around $y_{0}=0.6$ than WS case. Meanwhile in Fig.~\ref{fig:Fig4_ratio_fitting}, FFG and HMT show larger ratio in peak and ridge region with smaller ratio in marginal region. Comparison of $c_{n}$ between WS, FFG and HMT proved that, asymmetry from FFG and HMT at initial state transformed into different ratio at final state. Intrinsic momentum from FFG and HMT generates more events with larger $y_{0}$ in peak and ridge, corresponding to larger width of $y_{0}$, but the intrinsic momentum can not support larger $y_{0}$ to extend to edge around $y_{0}=0.6$. Then the asymmetry transformed into rapidity asymmetry in Fig.~\ref{fig:Fig3_rapidity_3_region_pdf} and Fig.~\ref{fig:Fig4_ratio_fitting}, intrinsic momentum from FFG and HMT provides us enhanced ratio in peak and depressed ratio in ridge and margin. That's why we see larger $c_{1}$ and smaller $c_{2},c_{3}$ in FFG and HMT than WS. 

{To test our interpretation, we also extract skewness and kurtosis of rapidity distribution in different systems and $\alpha_{ZN}$ regions in Fig.~\ref{fig:Fig6_skewness_kurtosis_new_50bin}.
According to knowledge from statistics, the skewness is defined as $\frac{\mu_3}{\sigma^3}$, and the kurtosis is defined as $\frac{\mu_4}{\sigma^4}-3$, in which $\mu_{n}=\langle (X-\langle X\rangle)^{n}\rangle$ and $\sigma = \sqrt{\frac{\Sigma (X_i-\langle X\rangle)^2}{N}}$. The skewness describes how far the events distribution deviates from the mean value, for a standard Gaussian distribution the skewness is 0, and a positive skewness shows a longer small tail on the right of the mean value because a few events with higher X distribute on the right. And the kurtosis describes how the events concentrate around the mean value, for a standard Gaussian distribution the kurtosis is also 0, and higher kurtosis means more events distribute around the mean value. The values of skewness in four different configurations (W-S, Tri, FFG, HMT) do not show significant dependence, but in same configuration we can always see the skewness in positive region is smaller than the one in middle region, then both smaller than the one in negative region, the relationship is consistent to the physical picture of rapidity shift. And the central values of kurtosis in FFG and HMT cases show a rising trend than Woods-Saxon cases as baseline. According to the statistical significance of kurtosis, higher kurtosis means more events concentrate to distribute around the mean value, that's to say, FFG and HMT cases have more events around the peak and less events in ridge and margin. This deformation of rapidity distribution leads to the anomaly of ratio and $c_{n}$. And further, comparing to the HMT case, larger kurtosis in positive region and smaller kurtosis in negative region in FFG case means in peak region, more events distribute in $+asym$ and less events distribute in $-syem$, which leads to higher $c_{1}$ in FFG than HMT. Here we see that the kurtosis also supports our physical picture for the effect on longitudinal asymmetry from intrinsic momentum distribution.}

Lastly, we can discuss the difference between FFG and HMT. According to Fig.~\ref{fig:Fig5_cnyn}, actually we can see the fitting line of $ratio_{+/-}$ of FFG (green) is higher than HMT (blue) in most region of peak and ridge as we mentioned in Fig.~\ref{fig:Fig3_rapidity_3_region_pdf}. Considering that $c_{1}$ dominates $ratio_{+/-}$ as shown in Fig.~\ref{fig:Fig5_cnyn}, we can say the effect of deformation in FFG is mainly shown as generating more events in peak of rapidity distribution and less events at the edge close to $\pm5$. It is reasonable for FFG indeed provides additional momentum distribution on $y_{0}$, because there is no interaction between nucleon, but meanwhile FFG can not provide more particles emitted to larger rapidity ($y\sim5$). To compensate the over-increasing $c_{1}$ which dominates in mid-rapidity region, $c_{2}$ and $c_{3}$ are both small enough to 0, even negative as shown in Fig.~\ref{fig:Fig5_cnyn} and Fig.~\ref{fig:Fig8_cn_tab}. But the SRC mechanism in HMT provides a way to emit more particles with larger rapidity. According to Ref.~\cite{CiofidegliAtti:1991mm,Wang:2017odj,Patsyuk:2021fju}, HMT can cause more high energy nucleon emission at final state, so in beam direction more particles with larger rapidity can distribute close to $\pm5$. That's why Fig.~\ref{fig:Fig8_cn_tab} and Fig.~\ref{fig:Fig5_cnyn} show us that $c_{2},c_{3}$ of HMT provide larger and positive contribution than those of FFG. In summary, intrinsic momentum distribution are transformed to different deformation of final rapidity distribution, and their effect on longitudinal asymmetry can be characterized by $c_{n}$.

\subsection{Prospect and alternative improvement in experiments}

 For both initial condition and longitudinal asymmetry we introduced above, some experiments have been carried out to investigate them prospectively. { Considering different progress in various experiments, here we can simply show some feasibility on our suggested future experiments. The SRC experiments on JLab extend to systems with larger removal energy, momentum and more complex nuclei which can give us insights about effect from HMT. And measurements of rapidity distribution or longitudinal asymmetry in heavy ion collisions have been proceeded like Au+Au or Pb+Pb on RHIC and LHC. Also }some are planned to investigate them
at FRIB and FAIR etc \cite{201820,Patsyuk:2021fju}. It's possible to test some joint measurements, for an instance, electron-nucleus scattering \cite{eA} experiments can help us estimate HMT component and short-range-correlation effect in ion collision \cite{CiofidegliAtti:1991mm,Patsyuk:2021fju}, meanwhile collective flow $v_{n}$, characteristic spectra of giant dipole resonance (GDR), dihadron azimuthal correlation and backward-forward multiplicity correlation can help us to distinguish $\alpha$-cluster structure \cite{He:2014iqa,Li:2020vrg,Li:2021znq,Li:2022bpm,Zhang:2021phk,Alver:2010gr,Guo:2017tco,Ma_NuclTech}, lastly the energy deposition in detector and rapidity measurement reveal the longitudinal asymmetry \cite{201820}. {Also, for some practical application, we suggest to measure rapidity spectrum or comparing previous data, and use probes of α-cluster or HMT mentioned above in RHIC or ALICE experiments to distinguish different configurations,  then measure coefficients $\langle y_{0}\rangle$ and $c_{n}$ in different configurations to match their longitudinal asymmetry. Then according to our work one can give these systems' initial rapidity shift and non-zero momentum shift, which can be used to reconstruct colliding vertex or correct the initial angular momentum, this reconstruction and correction may affect initial condition in polarization problem, chiral magnetic effect measurement and so on, so we hope our work on longitudinal asymmetry can be applied in observable measurements in these experiments.} {Despite progress of experimental program limited, RHIC and ALICE have provided us abundant benchmarks for the rapidity distribution and longitudinal asymmetry of various systems. These results can benefit future measurements that may be performed in more experiments on different colliders and systems such as O+O colliding on FRIB, SRC experiments on JLab and so on.} By carrying out these experiments in symmetric nuclei collision, we can give insight or provide improvement of physical picture on longitudinal asymmetry, further to constrain condition of collision and describe final rapidity more precisely.

\section{Summary}
\label{sec:summary}
This paper presents a comparison of the longitudinal asymmetry for systems with different $\alpha$ cluster structure and intrinsic momentum distribution in AMPT model. $\alpha_{ZN}$ and $y_{0}$ are calculated to characterise the rapidity shift, as performed in experimental measurements by ALICE \cite{201820}. To study the effect of different initial condition on longitudinal asymmetry, we introduce $\alpha$ cluster structure and different intrinsic nucleon momentum distribution into the simulation {in AMPT model}, where the intrinsic momentum distribution is added to the parameter $y_{0}$ as shown in Fig.~\ref{fig:Fig1_y0_alpha_ZN}(a3) and (b3). With these data we use 3rd polynomial fitting to extract the expansion coefficients $c_{n}$ in Fig.~\ref{fig:Fig8_cn_tab}. The comparison among different initial conditions shows us the effects of the $\alpha$ clustering structure and the initial momentum component.

\par Based on our analysis, we propose that the dependence of the longitudinal asymmetry is the result of the competition between rapidity shift and rapidity deformation. In the $|y| \textless 1$ region, $c_{1}$ is mainly linearly dependent on the initial rapidity shift if we don't consider the momentum distribution, and the momentum distribution will lead to rapidity deformation, shown as a larger ratio in peak and ridge. However, in the large rapidity region, $c_{2}$ and $c_{3}$ reflect the deformation of the final-state rapidity distribution. HMT caused by SRC provides a larger rapidity distribution when $y$ is close to $\pm 5$, which enhances the longitudinal asymmetry of $c_{2}$ and $c_{3}$. {To further support our interpretation, we extracted the skewness and kurtosis from different configurations. By comparing skewness and kurtosis from different shift regions and configurations, we demonstrate that the particle rapidity distribution caused by different configurations, especially FFG and HMT, provide deformation in different regions as expected, resulting in additional longitudinal asymmetry. And this phenomenon shows consistency with the anomaly of coefficients $c_{n}$, which can be measured in future experiments.}

\par Finally, we discuss practical application of our calculation in experiments, including joint measurement on $\alpha$-clustering effect, high momentum component effect, and longitudinal asymmetry with deformation, some dependent experiments have been performed in different detectors~\cite{201820,Patsyuk:2021fju,Guo:2017tco,Ma:2022dbh}. {And we respectively introduced
different experiments for the joint measurement, so that
researchers can ensure the feasibility of suggested experiments in the future.} In order to test the results of this work, we propose to investigate the collision of symmetric nuclei of the C + C system, and in the future we expect that these investigations{ by observable related to initial condition can provide us with insights to constrain the nuclear structure and the intrinsic momentum distribution of nucleon in the nuclear, as well as the correction for the deformation of the final rapidity distribution.} 

\begin{acknowledgements}

{This work was supported in part by National Key R\&D Program of China under Grant No. 2018YFE0104600, the National Natural Science Foundation of China under contract Nos.  11890710, 11890714, 12147101, 12275054, 11875066, 11925502, 11961141003 and the Strategic Priority Research Program of CAS under Grant No. XDB34000000, Guangdong Major Project of Basic and Applied Basic Research No. 2020B0301030008, and Shanghai Special Project for Basic Research No. 22TQ006.}

\end{acknowledgements}

	\end{CJK*}	

\bibliography{Longitude_Asymmetric}

\begin{thebibliography}{72}%
\makeatletter
\providecommand \@ifxundefined [1]{%
 \@ifx{#1\undefined}
}%
\providecommand \@ifnum [1]{%
 \ifnum #1\expandafter \@firstoftwo
 \else \expandafter \@secondoftwo
 \fi
}%
\providecommand \@ifx [1]{%
 \ifx #1\expandafter \@firstoftwo
 \else \expandafter \@secondoftwo
 \fi
}%
\providecommand \natexlab [1]{#1}%
\providecommand \enquote  [1]{``#1''}%
\providecommand \bibnamefont  [1]{#1}%
\providecommand \bibfnamefont [1]{#1}%
\providecommand \citenamefont [1]{#1}%
\providecommand \href@noop [0]{\@secondoftwo}%
\providecommand \href [0]{\begingroup \@sanitize@url \@href}%
\providecommand \@href[1]{\@@startlink{#1}\@@href}%
\providecommand \@@href[1]{\endgroup#1\@@endlink}%
\providecommand \@sanitize@url [0]{\catcode `\\12\catcode `\$12\catcode
  `\&12\catcode `\#12\catcode `\^12\catcode `\_12\catcode `\%12\relax}%
\providecommand \@@startlink[1]{}%
\providecommand \@@endlink[0]{}%
\providecommand \url  [0]{\begingroup\@sanitize@url \@url }%
\providecommand \@url [1]{\endgroup\@href {#1}{\urlprefix }}%
\providecommand \urlprefix  [0]{URL }%
\providecommand \Eprint [0]{\href }%
\providecommand \doibase [0]{https://doi.org/}%
\providecommand \selectlanguage [0]{\@gobble}%
\providecommand \bibinfo  [0]{\@secondoftwo}%
\providecommand \bibfield  [0]{\@secondoftwo}%
\providecommand \translation [1]{[#1]}%
\providecommand \BibitemOpen [0]{}%
\providecommand \bibitemStop [0]{}%
\providecommand \bibitemNoStop [0]{.\EOS\space}%
\providecommand \EOS [0]{\spacefactor3000\relax}%
\providecommand \BibitemShut  [1]{\csname bibitem#1\endcsname}%
\let\auto@bib@innerbib\@empty
\bibitem [{\citenamefont {Connors}\ \emph {et~al.}(2018)\citenamefont
  {Connors}, \citenamefont {Nattrass}, \citenamefont {Reed},\ and\
  \citenamefont {Salur}}]{RMP}%
  \BibitemOpen
  \bibfield  {author} {\bibinfo {author} {\bibfnamefont {M.}~\bibnamefont
  {Connors}}, \bibinfo {author} {\bibfnamefont {C.}~\bibnamefont {Nattrass}},
  \bibinfo {author} {\bibfnamefont {R.}~\bibnamefont {Reed}},\ and\ \bibinfo
  {author} {\bibfnamefont {S.}~\bibnamefont {Salur}},\ }\href
  {https://doi.org/10.1103/RevModPhys.90.025005} {\bibfield  {journal}
  {\bibinfo  {journal} {Rev. Mod. Phys.}\ }\textbf {\bibinfo {volume} {90}},\
  \bibinfo {pages} {025005} (\bibinfo {year} {2018})}\BibitemShut {NoStop}%
\bibitem [{\citenamefont {Bzdak}\ \emph {et~al.}(2020)\citenamefont {Bzdak},
  \citenamefont {Esumi}, \citenamefont {Koch}, \citenamefont {Liao},
  \citenamefont {Stephanov},\ and\ \citenamefont {Xu}}]{PhysRep1}%
  \BibitemOpen
  \bibfield  {author} {\bibinfo {author} {\bibfnamefont {A.}~\bibnamefont
  {Bzdak}}, \bibinfo {author} {\bibfnamefont {S.}~\bibnamefont {Esumi}},
  \bibinfo {author} {\bibfnamefont {V.}~\bibnamefont {Koch}}, \bibinfo {author}
  {\bibfnamefont {J.~F.}\ \bibnamefont {Liao}}, \bibinfo {author}
  {\bibfnamefont {M.}~\bibnamefont {Stephanov}},\ and\ \bibinfo {author}
  {\bibfnamefont {N.}~\bibnamefont {Xu}},\ }\href
  {https://doi.org/10.1016/j.physrep.2020.01.005} {\bibfield  {journal}
  {\bibinfo  {journal} {Phys. Rep.}\ }\textbf {\bibinfo {volume} {853}},\
  \bibinfo {pages} {1} (\bibinfo {year} {2020})}\BibitemShut {NoStop}%
\bibitem [{\citenamefont {Rothkopf}(2020)}]{PhysRep2}%
  \BibitemOpen
  \bibfield  {author} {\bibinfo {author} {\bibfnamefont {A.}~\bibnamefont
  {Rothkopf}},\ }\href {https://doi.org/10.1016/j.physrep.2020.02.006}
  {\bibfield  {journal} {\bibinfo  {journal} {Phys. Rep.}\ }\textbf {\bibinfo
  {volume} {858}},\ \bibinfo {pages} {1} (\bibinfo {year} {2020})}\BibitemShut
  {NoStop}%
\bibitem [{\citenamefont {Chen}\ \emph {et~al.}(2018)\citenamefont {Chen},
  \citenamefont {Keane}, \citenamefont {Ma}, \citenamefont {Tang},\ and\
  \citenamefont {Xu}}]{PhysRep3}%
  \BibitemOpen
  \bibfield  {author} {\bibinfo {author} {\bibfnamefont {J.}~\bibnamefont
  {Chen}}, \bibinfo {author} {\bibfnamefont {D.}~\bibnamefont {Keane}},
  \bibinfo {author} {\bibfnamefont {Y.~G.}\ \bibnamefont {Ma}}, \bibinfo
  {author} {\bibfnamefont {A.}~\bibnamefont {Tang}},\ and\ \bibinfo {author}
  {\bibfnamefont {Z.}~\bibnamefont {Xu}},\ }\href
  {https://doi.org/10.1016/j.physrep.2018.07.002} {\bibfield  {journal}
  {\bibinfo  {journal} {Phys. Rep.}\ }\textbf {\bibinfo {volume} {760}},\
  \bibinfo {pages} {1} (\bibinfo {year} {2018})}\BibitemShut {NoStop}%
\bibitem [{\citenamefont {Abdulhamid}\ \emph {et~al.}(2023)\citenamefont
  {Abdulhamid} \emph {et~al.}}]{PRL1}%
  \BibitemOpen
  \bibfield  {author} {\bibinfo {author} {\bibfnamefont {M.~I.}\ \bibnamefont
  {Abdulhamid}} \emph {et~al.},\ }\href
  {https://doi.org/10.1103/PhysRevLett.130.202301} {\bibfield  {journal}
  {\bibinfo  {journal} {Phys. Rev. Lett.}\ }\textbf {\bibinfo {volume} {130}},\
  \bibinfo {pages} {202301} (\bibinfo {year} {2023})}\BibitemShut {NoStop}%
\bibitem [{\citenamefont {Tumasyan}\ \emph {et~al.}(2022)\citenamefont
  {Tumasyan} \emph {et~al.}}]{PRL2}%
  \BibitemOpen
  \bibfield  {author} {\bibinfo {author} {\bibfnamefont {A.}~\bibnamefont
  {Tumasyan}} \emph {et~al.},\ }\href
  {https://doi.org/10.1103/PhysRevLett.129.022001} {\bibfield  {journal}
  {\bibinfo  {journal} {Phys. Rev. Lett.}\ }\textbf {\bibinfo {volume} {129}},\
  \bibinfo {pages} {022001} (\bibinfo {year} {2022})}\BibitemShut {NoStop}%
\bibitem [{\citenamefont {Zhu}\ \emph {et~al.}(2022)\citenamefont {Zhu},
  \citenamefont {Wang}, \citenamefont {Wang},\ and\ \citenamefont
  {Zheng}}]{ZhuLL}%
  \BibitemOpen
  \bibfield  {author} {\bibinfo {author} {\bibfnamefont {L.~L.}\ \bibnamefont
  {Zhu}}, \bibinfo {author} {\bibfnamefont {B.}~\bibnamefont {Wang}}, \bibinfo
  {author} {\bibfnamefont {M.}~\bibnamefont {Wang}},\ and\ \bibinfo {author}
  {\bibfnamefont {H.}~\bibnamefont {Zheng}},\ }\href
  {https://doi.org/10.1007/s41365-022-01028-8} {\bibfield  {journal} {\bibinfo
  {journal} {Nucl. Sci. Tech.}\ }\textbf {\bibinfo {volume} {33}},\ \bibinfo
  {pages} {45} (\bibinfo {year} {2022})}\BibitemShut {NoStop}%
\bibitem [{\citenamefont {Wang}\ \emph {et~al.}(2022)\citenamefont {Wang},
  \citenamefont {Tao}, \citenamefont {Zheng}, \citenamefont {Zhang},
  \citenamefont {Zhu},\ and\ \citenamefont {Bonasera}}]{ZhuLL2}%
  \BibitemOpen
  \bibfield  {author} {\bibinfo {author} {\bibfnamefont {M.}~\bibnamefont
  {Wang}}, \bibinfo {author} {\bibfnamefont {J.~Q.}\ \bibnamefont {Tao}},
  \bibinfo {author} {\bibfnamefont {H.}~\bibnamefont {Zheng}}, \bibinfo
  {author} {\bibfnamefont {W.~C.}\ \bibnamefont {Zhang}}, \bibinfo {author}
  {\bibfnamefont {L.~L.}\ \bibnamefont {Zhu}},\ and\ \bibinfo {author}
  {\bibfnamefont {A.}~\bibnamefont {Bonasera}},\ }\href
  {https://doi.org/10.1007/s41365-022-01019-9} {\bibfield  {journal} {\bibinfo
  {journal} {Nucl. Sci. Tech.}\ }\textbf {\bibinfo {volume} {33}},\ \bibinfo
  {pages} {37} (\bibinfo {year} {2022})}\BibitemShut {NoStop}%
\bibitem [{\citenamefont {Lan}\ and\ \citenamefont {Shi}(2022)}]{SSS}%
  \BibitemOpen
  \bibfield  {author} {\bibinfo {author} {\bibfnamefont {S.~W.}\ \bibnamefont
  {Lan}}\ and\ \bibinfo {author} {\bibfnamefont {S.~S.}\ \bibnamefont {Shi}},\
  }\href {https://doi.org/10.1007/s41365-022-01006-0} {\bibfield  {journal}
  {\bibinfo  {journal} {Nucl. Sci. Tech.}\ }\textbf {\bibinfo {volume} {33}},\
  \bibinfo {pages} {21} (\bibinfo {year} {2022})}\BibitemShut {NoStop}%
\bibitem [{\citenamefont {Liu}\ and\ \citenamefont {Huang}(2022)}]{Huang}%
  \BibitemOpen
  \bibfield  {author} {\bibinfo {author} {\bibfnamefont {Y.~C.}\ \bibnamefont
  {Liu}}\ and\ \bibinfo {author} {\bibfnamefont {X.~G.}\ \bibnamefont
  {Huang}},\ }\href {https://doi.org/10.1007/s11433-022-1903-8} {\bibfield
  {journal} {\bibinfo  {journal} {Science China - Phys. Mech. Astro.}\ }\textbf
  {\bibinfo {volume} {65}},\ \bibinfo {pages} {272011} (\bibinfo {year}
  {2022})}\BibitemShut {NoStop}%
\bibitem [{\citenamefont {Zhang}\ \emph {et~al.}(2023)\citenamefont {Zhang},
  \citenamefont {Zhang},\ and\ \citenamefont {Luo}}]{Luo}%
  \BibitemOpen
  \bibfield  {author} {\bibinfo {author} {\bibfnamefont {Y.}~\bibnamefont
  {Zhang}}, \bibinfo {author} {\bibfnamefont {D.}~\bibnamefont {Zhang}},\ and\
  \bibinfo {author} {\bibfnamefont {X.}~\bibnamefont {Luo}},\ }\href
  {https://doi.org/10.11889/j.0253-3219.2023.hjs.46.040001} {\bibfield
  {journal} {\bibinfo  {journal} {Nucl. Tech. (in Chinese)}\ }\textbf {\bibinfo
  {volume} {46}},\ \bibinfo {pages} {040001} (\bibinfo {year}
  {2023})}\BibitemShut {NoStop}%
\bibitem [{\citenamefont {Ko}(2023)}]{Ko}%
  \BibitemOpen
  \bibfield  {author} {\bibinfo {author} {\bibfnamefont {C.~M.}\ \bibnamefont
  {Ko}},\ }\href {https://doi.org/https://doi.org/10.1007/s41365-023-01231-1}
  {\bibfield  {journal} {\bibinfo  {journal} {Nucl. Sci. Tech.}\ }\textbf
  {\bibinfo {volume} {34}},\ \bibinfo {pages} {80} (\bibinfo {year}
  {2023})}\BibitemShut {NoStop}%
\bibitem [{\citenamefont {Sun}\ \emph {et~al.}(2023)\citenamefont {Sun},
  \citenamefont {Chen}, \citenamefont {Ko}, \citenamefont {Li}, \citenamefont
  {Xu},\ and\ \citenamefont {Xu}}]{Sun}%
  \BibitemOpen
  \bibfield  {author} {\bibinfo {author} {\bibfnamefont {K.}~\bibnamefont
  {Sun}}, \bibinfo {author} {\bibfnamefont {L.}~\bibnamefont {Chen}}, \bibinfo
  {author} {\bibfnamefont {C.~M.}\ \bibnamefont {Ko}}, \bibinfo {author}
  {\bibfnamefont {F.}~\bibnamefont {Li}}, \bibinfo {author} {\bibfnamefont
  {J.}~\bibnamefont {Xu}},\ and\ \bibinfo {author} {\bibfnamefont
  {Z.}~\bibnamefont {Xu}},\ }\href
  {https://doi.org/10.11889/j.0253-3219.2023.hjs.46.040012} {\bibfield
  {journal} {\bibinfo  {journal} {Nucl. Tech. (in Chinese)}\ }\textbf {\bibinfo
  {volume} {46}},\ \bibinfo {pages} {040012} (\bibinfo {year}
  {2023})}\BibitemShut {NoStop}%
\bibitem [{\citenamefont {Rapp}(2023)}]{Rapp}%
  \BibitemOpen
  \bibfield  {author} {\bibinfo {author} {\bibfnamefont {R.}~\bibnamefont
  {Rapp}},\ }\href {https://doi.org/10.1007/s41365-023-01213-3} {\bibfield
  {journal} {\bibinfo  {journal} {Nucl. Sci. Tech.}\ }\textbf {\bibinfo
  {volume} {34}},\ \bibinfo {pages} {63} (\bibinfo {year} {2023})}\BibitemShut
  {NoStop}%
\bibitem [{\citenamefont {Thakur}\ \emph {et~al.}(2022)\citenamefont {Thakur},
  \citenamefont {Saha~Sumit} \emph {et~al.}}]{longi_Thakur}%
  \BibitemOpen
  \bibfield  {author} {\bibinfo {author} {\bibfnamefont {S.}~\bibnamefont
  {Thakur}}, \bibinfo {author} {\bibfnamefont {K.}~\bibnamefont {Saha~Sumit}},
  \emph {et~al.},\ }\href {https://doi.org/10.1140/epja/s10050-022-00667-0}
  {\bibfield  {journal} {\bibinfo  {journal} {The European Physical Journal A}\
  }\textbf {\bibinfo {volume} {58}},\ \bibinfo {pages} {13} (\bibinfo {year}
  {2022})}\BibitemShut {NoStop}%
\bibitem [{\citenamefont {Raniwala}\ \emph {et~al.}(2018)\citenamefont
  {Raniwala}, \citenamefont {Raniwala},\ and\ \citenamefont
  {Loizides}}]{PhysRevC.97.024912}%
  \BibitemOpen
  \bibfield  {author} {\bibinfo {author} {\bibfnamefont {R.}~\bibnamefont
  {Raniwala}}, \bibinfo {author} {\bibfnamefont {S.}~\bibnamefont {Raniwala}},\
  and\ \bibinfo {author} {\bibfnamefont {C.}~\bibnamefont {Loizides}},\ }\href
  {https://doi.org/10.1103/PhysRevC.97.024912} {\bibfield  {journal} {\bibinfo
  {journal} {Phys. Rev. C}\ }\textbf {\bibinfo {volume} {97}},\ \bibinfo
  {pages} {024912} (\bibinfo {year} {2018})}\BibitemShut {NoStop}%
\bibitem [{\citenamefont {Deng}\ \emph {et~al.}(2022)\citenamefont {Deng},
  \citenamefont {Huang},\ and\ \citenamefont {Ma}}]{Deng:2021miw}%
  \BibitemOpen
  \bibfield  {author} {\bibinfo {author} {\bibfnamefont {X.-G.}\ \bibnamefont
  {Deng}}, \bibinfo {author} {\bibfnamefont {X.-G.}\ \bibnamefont {Huang}},\
  and\ \bibinfo {author} {\bibfnamefont {Y.-G.}\ \bibnamefont {Ma}},\ }\href
  {https://doi.org/10.1016/j.physletb.2022.137560} {\bibfield  {journal}
  {\bibinfo  {journal} {Phys. Lett. B}\ }\textbf {\bibinfo {volume} {835}},\
  \bibinfo {pages} {137560} (\bibinfo {year} {2022})}\BibitemShut {NoStop}%
\bibitem [{\citenamefont {Bleicher}\ \emph {et~al.}(1999)\citenamefont
  {Bleicher}, \citenamefont {Zabrodin}, \citenamefont {Spieles} \emph
  {et~al.}}]{UrQMD}%
  \BibitemOpen
  \bibfield  {author} {\bibinfo {author} {\bibfnamefont {M.}~\bibnamefont
  {Bleicher}}, \bibinfo {author} {\bibfnamefont {E.}~\bibnamefont {Zabrodin}},
  \bibinfo {author} {\bibfnamefont {C.}~\bibnamefont {Spieles}}, \emph
  {et~al.},\ }\href {https://doi.org/10.1088/0954-3899/25/9/308} {\bibfield
  {journal} {\bibinfo  {journal} {J. Phys. G}\ }\textbf {\bibinfo {volume}
  {25}},\ \bibinfo {pages} {1859} (\bibinfo {year} {1999})}\BibitemShut
  {NoStop}%
\bibitem [{\citenamefont {Xiao}\ \emph {et~al.}(2023)\citenamefont {Xiao},
  \citenamefont {Li}, \citenamefont {Wang}, \citenamefont {Liu},\ and\
  \citenamefont {Li}}]{UrQMD1}%
  \BibitemOpen
  \bibfield  {author} {\bibinfo {author} {\bibfnamefont {K.}~\bibnamefont
  {Xiao}}, \bibinfo {author} {\bibfnamefont {P.~C.}\ \bibnamefont {Li}},
  \bibinfo {author} {\bibfnamefont {Y.~J.}\ \bibnamefont {Wang}}, \bibinfo
  {author} {\bibfnamefont {F.~H.}\ \bibnamefont {Liu}},\ and\ \bibinfo {author}
  {\bibfnamefont {Q.~F.}\ \bibnamefont {Li}},\ }\href
  {https://doi.org/10.1007/s41365-023-01205-3} {\bibfield  {journal} {\bibinfo
  {journal} {Nucl. Sci. Tech.}\ }\textbf {\bibinfo {volume} {34}},\ \bibinfo
  {pages} {62} (\bibinfo {year} {2023})}\BibitemShut {NoStop}%
\bibitem [{\citenamefont {Li}\ \emph {et~al.}(2023)\citenamefont {Li},
  \citenamefont {Steinheimer}, \citenamefont {Reichert} \emph
  {et~al.}}]{UrQMD2}%
  \BibitemOpen
  \bibfield  {author} {\bibinfo {author} {\bibfnamefont {P.~C.}\ \bibnamefont
  {Li}}, \bibinfo {author} {\bibfnamefont {J.}~\bibnamefont {Steinheimer}},
  \bibinfo {author} {\bibfnamefont {T.}~\bibnamefont {Reichert}}, \emph
  {et~al.},\ }\href {https://doi.org/10.1007/s11433-022-2041-8} {\bibfield
  {journal} {\bibinfo  {journal} {Science China - Phys. Mech. Astro.}\ }\textbf
  {\bibinfo {volume} {66}},\ \bibinfo {pages} {232011} (\bibinfo {year}
  {2023})}\BibitemShut {NoStop}%
\bibitem [{\citenamefont {Xi}\ \emph {et~al.}(2023)\citenamefont {Xi},
  \citenamefont {Deng}, \citenamefont {Zhang},\ and\ \citenamefont {Ma}}]{Xi}%
  \BibitemOpen
  \bibfield  {author} {\bibinfo {author} {\bibfnamefont {B.-S.}\ \bibnamefont
  {Xi}}, \bibinfo {author} {\bibfnamefont {X.-G.}\ \bibnamefont {Deng}},
  \bibinfo {author} {\bibfnamefont {S.}~\bibnamefont {Zhang}},\ and\ \bibinfo
  {author} {\bibfnamefont {Y.-G.}\ \bibnamefont {Ma}},\ }\href
  {https://doi.org/10.1140/epja/s10050-023-00932-w} {\bibfield  {journal}
  {\bibinfo  {journal} {Euro. Phys. J. A}\ }\textbf {\bibinfo {volume} {59}},\
  \bibinfo {pages} {33} (\bibinfo {year} {2023})}\BibitemShut {NoStop}%
\bibitem [{\citenamefont {Kornas}(2021)}]{HADES}%
  \BibitemOpen
  \bibfield  {author} {\bibinfo {author} {\bibfnamefont {F.}~\bibnamefont
  {Kornas}} (\bibinfo {collaboration} {for HADES Collaboration}),\ }\href@noop
  {} {} (\bibinfo {year} {2021}),\ \bibinfo {note} {talk given at Strangeness
  Quark Matter 2021 Online}\BibitemShut {NoStop}%
\bibitem [{\citenamefont {Abdallah}\ \emph {et~al.}(2021)\citenamefont
  {Abdallah} \emph {et~al.}}]{PhysRevC.104.L061901}%
  \BibitemOpen
  \bibfield  {author} {\bibinfo {author} {\bibfnamefont {M.~S.}\ \bibnamefont
  {Abdallah}} \emph {et~al.} (\bibinfo {collaboration} {STAR Collaboration}),\
  }\href {https://doi.org/10.1103/PhysRevC.104.L061901} {\bibfield  {journal}
  {\bibinfo  {journal} {Phys. Rev. C}\ }\textbf {\bibinfo {volume} {104}},\
  \bibinfo {pages} {L061901} (\bibinfo {year} {2021})}\BibitemShut {NoStop}%
\bibitem [{\citenamefont {Gamow}\ and\ \citenamefont
  {Rutherford}(1930)}]{article_Gamow}%
  \BibitemOpen
  \bibfield  {author} {\bibinfo {author} {\bibfnamefont {G.}~\bibnamefont
  {Gamow}}\ and\ \bibinfo {author} {\bibfnamefont {E.}~\bibnamefont
  {Rutherford}},\ }\href {https://doi.org/10.1098/rspa.1930.0032} {\bibfield
  {journal} {\bibinfo  {journal} {Proceedings of the Royal Society of London.
  Series A, Containing Papers of a Mathematical and Physical Character}\
  }\textbf {\bibinfo {volume} {126}},\ \bibinfo {pages} {632} (\bibinfo {year}
  {1930})}\BibitemShut {NoStop}%
\bibitem [{\citenamefont {Qin}\ \emph {et~al.}(2012)\citenamefont {Qin},
  \citenamefont {Hagel}, \citenamefont {Wada} \emph {et~al.}}]{Qin}%
  \BibitemOpen
  \bibfield  {author} {\bibinfo {author} {\bibfnamefont {L.}~\bibnamefont
  {Qin}}, \bibinfo {author} {\bibfnamefont {K.}~\bibnamefont {Hagel}}, \bibinfo
  {author} {\bibfnamefont {R.}~\bibnamefont {Wada}}, \emph {et~al.},\ }\href
  {https://doi.org/10.1103/PhysRevLett.108.172701} {\bibfield  {journal}
  {\bibinfo  {journal} {Phys. Rev. Lett.}\ }\textbf {\bibinfo {volume} {108}},\
  \bibinfo {pages} {172701} (\bibinfo {year} {2012})}\BibitemShut {NoStop}%
\bibitem [{\citenamefont {He}\ \emph {et~al.}(2014)\citenamefont {He},
  \citenamefont {Ma}, \citenamefont {Cao}, \citenamefont {Cai},\ and\
  \citenamefont {Zhang}}]{He:2014iqa}%
  \BibitemOpen
  \bibfield  {author} {\bibinfo {author} {\bibfnamefont {W.~B.}\ \bibnamefont
  {He}}, \bibinfo {author} {\bibfnamefont {Y.~G.}\ \bibnamefont {Ma}}, \bibinfo
  {author} {\bibfnamefont {X.~G.}\ \bibnamefont {Cao}}, \bibinfo {author}
  {\bibfnamefont {X.~Z.}\ \bibnamefont {Cai}},\ and\ \bibinfo {author}
  {\bibfnamefont {G.~Q.}\ \bibnamefont {Zhang}},\ }\href
  {https://doi.org/10.1103/PhysRevLett.113.032506} {\bibfield  {journal}
  {\bibinfo  {journal} {Phys. Rev. Lett.}\ }\textbf {\bibinfo {volume} {113}},\
  \bibinfo {pages} {032506} (\bibinfo {year} {2014})}\BibitemShut {NoStop}%
\bibitem [{\citenamefont {He}\ \emph {et~al.}(2023{\natexlab{a}})\citenamefont
  {He}, \citenamefont {Li}, \citenamefont {Ma}, \citenamefont {Niu},
  \citenamefont {Pei},\ and\ \citenamefont {Zhang}}]{Ma_ML}%
  \BibitemOpen
  \bibfield  {author} {\bibinfo {author} {\bibfnamefont {W.-B.}\ \bibnamefont
  {He}}, \bibinfo {author} {\bibfnamefont {Q.-F.}\ \bibnamefont {Li}}, \bibinfo
  {author} {\bibfnamefont {Y.-G.}\ \bibnamefont {Ma}}, \bibinfo {author}
  {\bibfnamefont {Z.-M.}\ \bibnamefont {Niu}}, \bibinfo {author} {\bibfnamefont
  {J.-C.}\ \bibnamefont {Pei}},\ and\ \bibinfo {author} {\bibfnamefont {Y.-X.}\
  \bibnamefont {Zhang}},\ }\href {https://doi.org/10.1007/s11433-023-2116-0}
  {\bibfield  {journal} {\bibinfo  {journal} {Science China - Phys. Mech.
  Astro.}\ }\textbf {\bibinfo {volume} {66}},\ \bibinfo {pages} {282001}
  (\bibinfo {year} {2023}{\natexlab{a}})}\BibitemShut {NoStop}%
\bibitem [{\citenamefont {He}\ \emph {et~al.}(2023{\natexlab{b}})\citenamefont
  {He}, \citenamefont {Ma}, \citenamefont {Pang}, \citenamefont {Song},\ and\
  \citenamefont {Zhou}}]{Ma_NST}%
  \BibitemOpen
  \bibfield  {author} {\bibinfo {author} {\bibfnamefont {W.}~\bibnamefont
  {He}}, \bibinfo {author} {\bibfnamefont {Y.}~\bibnamefont {Ma}}, \bibinfo
  {author} {\bibfnamefont {L.}~\bibnamefont {Pang}}, \bibinfo {author}
  {\bibfnamefont {H.}~\bibnamefont {Song}},\ and\ \bibinfo {author}
  {\bibfnamefont {K.}~\bibnamefont {Zhou}},\ }\href
  {https://doi.org/10.1007/s41365-023-01233-z} {\bibfield  {journal} {\bibinfo
  {journal} {Nucl. Sci. Tech.}\ }\textbf {\bibinfo {volume} {34}},\ \bibinfo
  {pages} {88} (\bibinfo {year} {2023}{\natexlab{b}})}\BibitemShut {NoStop}%
\bibitem [{\citenamefont {Broniowski}\ and\ \citenamefont
  {Ruiz~Arriola}(2014)}]{PhysRevLett.112.112501}%
  \BibitemOpen
  \bibfield  {author} {\bibinfo {author} {\bibfnamefont {W.}~\bibnamefont
  {Broniowski}}\ and\ \bibinfo {author} {\bibfnamefont {E.}~\bibnamefont
  {Ruiz~Arriola}},\ }\href {https://doi.org/10.1103/PhysRevLett.112.112501}
  {\bibfield  {journal} {\bibinfo  {journal} {Phys. Rev. Lett.}\ }\textbf
  {\bibinfo {volume} {112}},\ \bibinfo {pages} {112501} (\bibinfo {year}
  {2014})}\BibitemShut {NoStop}%
\bibitem [{\citenamefont {Li}\ \emph {et~al.}(2020)\citenamefont {Li},
  \citenamefont {Zhang},\ and\ \citenamefont {Ma}}]{Li:2020vrg}%
  \BibitemOpen
  \bibfield  {author} {\bibinfo {author} {\bibfnamefont {Y.-A.}\ \bibnamefont
  {Li}}, \bibinfo {author} {\bibfnamefont {S.}~\bibnamefont {Zhang}},\ and\
  \bibinfo {author} {\bibfnamefont {Y.-G.}\ \bibnamefont {Ma}},\ }\href
  {https://doi.org/10.1103/PhysRevC.102.054907} {\bibfield  {journal} {\bibinfo
   {journal} {Phys. Rev. C}\ }\textbf {\bibinfo {volume} {102}},\ \bibinfo
  {pages} {054907} (\bibinfo {year} {2020})}\BibitemShut {NoStop}%
\bibitem [{\citenamefont {Guo}\ \emph {et~al.}(2017)\citenamefont {Guo},
  \citenamefont {He},\ and\ \citenamefont {Ma}}]{Guo:2017tco}%
  \BibitemOpen
  \bibfield  {author} {\bibinfo {author} {\bibfnamefont {C.-C.}\ \bibnamefont
  {Guo}}, \bibinfo {author} {\bibfnamefont {W.-B.}\ \bibnamefont {He}},\ and\
  \bibinfo {author} {\bibfnamefont {Y.-G.}\ \bibnamefont {Ma}},\ }\href
  {https://doi.org/10.1088/0256-307X/34/9/092101} {\bibfield  {journal}
  {\bibinfo  {journal} {Chin. Phys. Lett.}\ }\textbf {\bibinfo {volume} {34}},\
  \bibinfo {pages} {092101} (\bibinfo {year} {2017})}\BibitemShut {NoStop}%
\bibitem [{\citenamefont {Li}\ \emph {et~al.}(2022{\natexlab{a}})\citenamefont
  {Li}, \citenamefont {Wang}, \citenamefont {Zhang},\ and\ \citenamefont
  {Ma}}]{Li:2022bpm}%
  \BibitemOpen
  \bibfield  {author} {\bibinfo {author} {\bibfnamefont {Y.-A.}\ \bibnamefont
  {Li}}, \bibinfo {author} {\bibfnamefont {D.-F.}\ \bibnamefont {Wang}},
  \bibinfo {author} {\bibfnamefont {S.}~\bibnamefont {Zhang}},\ and\ \bibinfo
  {author} {\bibfnamefont {Y.-G.}\ \bibnamefont {Ma}},\ }\href
  {https://doi.org/10.1088/1674-1137/ac3bc9} {\bibfield  {journal} {\bibinfo
  {journal} {Chin. Phys. C}\ }\textbf {\bibinfo {volume} {46}},\ \bibinfo
  {pages} {044101} (\bibinfo {year} {2022}{\natexlab{a}})}\BibitemShut
  {NoStop}%
\bibitem [{\citenamefont {Li}\ \emph {et~al.}(2021)\citenamefont {Li},
  \citenamefont {Wang}, \citenamefont {Zhang},\ and\ \citenamefont
  {Ma}}]{Li:2021znq}%
  \BibitemOpen
  \bibfield  {author} {\bibinfo {author} {\bibfnamefont {Y.-A.}\ \bibnamefont
  {Li}}, \bibinfo {author} {\bibfnamefont {D.-F.}\ \bibnamefont {Wang}},
  \bibinfo {author} {\bibfnamefont {S.}~\bibnamefont {Zhang}},\ and\ \bibinfo
  {author} {\bibfnamefont {Y.-G.}\ \bibnamefont {Ma}},\ }\href
  {https://doi.org/10.1103/PhysRevC.104.044906} {\bibfield  {journal} {\bibinfo
   {journal} {Phys. Rev. C}\ }\textbf {\bibinfo {volume} {104}},\ \bibinfo
  {pages} {044906} (\bibinfo {year} {2021})}\BibitemShut {NoStop}%
\bibitem [{\citenamefont {Ma}\ and\ \citenamefont {Zhang}(2020)}]{Ma:2022dbh}%
  \BibitemOpen
  \bibfield  {author} {\bibinfo {author} {\bibfnamefont {Y.-G.}\ \bibnamefont
  {Ma}}\ and\ \bibinfo {author} {\bibfnamefont {S.}~\bibnamefont {Zhang}},\
  }\bibinfo {title} {Influence of nuclear structure in relativistic heavy-ion
  collisions},\ in\ \href {https://doi.org/10.1007/978-981-15-8818-1_5-1}
  {\emph {\bibinfo {booktitle} {Handbook of Nuclear Physics}}},\ \bibinfo
  {editor} {edited by\ \bibinfo {editor} {\bibfnamefont {I.}~\bibnamefont
  {Tanihata}}, \bibinfo {editor} {\bibfnamefont {H.}~\bibnamefont {Toki}},\
  and\ \bibinfo {editor} {\bibfnamefont {T.}~\bibnamefont {Kajino}}}\ (\bibinfo
   {publisher} {Springer Nature Singapore},\ \bibinfo {address} {Singapore},\
  \bibinfo {year} {2020})\ pp.\ \bibinfo {pages} {1--30}\BibitemShut {NoStop}%
\bibitem [{\citenamefont {Ma}(2023)}]{Ma_NuclTech}%
  \BibitemOpen
  \bibfield  {author} {\bibinfo {author} {\bibfnamefont {Y.-G.}\ \bibnamefont
  {Ma}},\ }\href {https://doi.org/10.11889/j.0253 3219.2023.hjs.46.080001}
  {\bibfield  {journal} {\bibinfo  {journal} {Nucl. Tech. (in Chinese)}\
  }\textbf {\bibinfo {volume} {46}},\ \bibinfo {pages} {080001} (\bibinfo
  {year} {2023})}\BibitemShut {NoStop}%
\bibitem [{\citenamefont {Antonov}\ \emph {et~al.}(1980)\citenamefont
  {Antonov}, \citenamefont {Nikolaev},\ and\ \citenamefont
  {Petkov}}]{Antonov1980}%
  \BibitemOpen
  \bibfield  {author} {\bibinfo {author} {\bibfnamefont {A.~N.}\ \bibnamefont
  {Antonov}}, \bibinfo {author} {\bibfnamefont {V.~A.}\ \bibnamefont
  {Nikolaev}},\ and\ \bibinfo {author} {\bibfnamefont {I.~Z.}\ \bibnamefont
  {Petkov}},\ }\href {https://doi.org/10.1007/BF01892806} {\bibfield  {journal}
  {\bibinfo  {journal} {Zeitschrift f\"ur Physik A Atoms and Nuclei}\ }\textbf
  {\bibinfo {volume} {297}},\ \bibinfo {pages} {257} (\bibinfo {year}
  {1980})}\BibitemShut {NoStop}%
\bibitem [{\citenamefont {Ciofi~degli Atti}\ \emph {et~al.}(1991)\citenamefont
  {Ciofi~degli Atti}, \citenamefont {Simula}, \citenamefont {Frankfurt},\ and\
  \citenamefont {Strikman}}]{CiofidegliAtti:1991mm}%
  \BibitemOpen
  \bibfield  {author} {\bibinfo {author} {\bibfnamefont {C.}~\bibnamefont
  {Ciofi~degli Atti}}, \bibinfo {author} {\bibfnamefont {S.}~\bibnamefont
  {Simula}}, \bibinfo {author} {\bibfnamefont {L.~L.}\ \bibnamefont
  {Frankfurt}},\ and\ \bibinfo {author} {\bibfnamefont {M.~I.}\ \bibnamefont
  {Strikman}},\ }\href {https://doi.org/10.1103/PhysRevC.44.R7} {\bibfield
  {journal} {\bibinfo  {journal} {Phys. Rev. C}\ }\textbf {\bibinfo {volume}
  {44}},\ \bibinfo {pages} {R7} (\bibinfo {year} {1991})}\BibitemShut {NoStop}%
\bibitem [{\citenamefont {Wang}\ \emph {et~al.}(2017)\citenamefont {Wang},
  \citenamefont {Xu}, \citenamefont {Ren},\ and\ \citenamefont
  {Gao}}]{Wang:2017odj}%
  \BibitemOpen
  \bibfield  {author} {\bibinfo {author} {\bibfnamefont {Z.}~\bibnamefont
  {Wang}}, \bibinfo {author} {\bibfnamefont {C.}~\bibnamefont {Xu}}, \bibinfo
  {author} {\bibfnamefont {Z.}~\bibnamefont {Ren}},\ and\ \bibinfo {author}
  {\bibfnamefont {C.}~\bibnamefont {Gao}},\ }\href
  {https://doi.org/10.1103/PhysRevC.96.054603} {\bibfield  {journal} {\bibinfo
  {journal} {Phys. Rev. C}\ }\textbf {\bibinfo {volume} {96}},\ \bibinfo
  {pages} {054603} (\bibinfo {year} {2017})}\BibitemShut {NoStop}%
\bibitem [{\citenamefont {Shen}\ \emph {et~al.}(2022)\citenamefont {Shen},
  \citenamefont {Huang},\ and\ \citenamefont {Ma}}]{Shen:2021dll}%
  \BibitemOpen
  \bibfield  {author} {\bibinfo {author} {\bibfnamefont {L.}~\bibnamefont
  {Shen}}, \bibinfo {author} {\bibfnamefont {B.-S.}\ \bibnamefont {Huang}},\
  and\ \bibinfo {author} {\bibfnamefont {Y.-G.}\ \bibnamefont {Ma}},\ }\href
  {https://doi.org/10.1103/PhysRevC.105.014603} {\bibfield  {journal} {\bibinfo
   {journal} {Phys. Rev. C}\ }\textbf {\bibinfo {volume} {105}},\ \bibinfo
  {pages} {014603} (\bibinfo {year} {2022})}\BibitemShut {NoStop}%
\bibitem [{\citenamefont {Lin}\ \emph {et~al.}(2005)\citenamefont {Lin},
  \citenamefont {Ko}, \citenamefont {Li}, \citenamefont {Zhang},\ and\
  \citenamefont {Pal}}]{AMPT2005}%
  \BibitemOpen
  \bibfield  {author} {\bibinfo {author} {\bibfnamefont {Z.-W.}\ \bibnamefont
  {Lin}}, \bibinfo {author} {\bibfnamefont {C.~M.}\ \bibnamefont {Ko}},
  \bibinfo {author} {\bibfnamefont {B.-A.}\ \bibnamefont {Li}}, \bibinfo
  {author} {\bibfnamefont {B.}~\bibnamefont {Zhang}},\ and\ \bibinfo {author}
  {\bibfnamefont {S.}~\bibnamefont {Pal}},\ }\href
  {https://doi.org/10.1103/PhysRevC.72.064901} {\bibfield  {journal} {\bibinfo
  {journal} {Phys. Rev. C}\ }\textbf {\bibinfo {volume} {72}},\ \bibinfo
  {pages} {064901} (\bibinfo {year} {2005})}\BibitemShut {NoStop}%
\bibitem [{\citenamefont {Ma}\ and\ \citenamefont {Lin}(2016)}]{AMPTGLM2016}%
  \BibitemOpen
  \bibfield  {author} {\bibinfo {author} {\bibfnamefont {G.-L.}\ \bibnamefont
  {Ma}}\ and\ \bibinfo {author} {\bibfnamefont {Z.-W.}\ \bibnamefont {Lin}},\
  }\href {https://doi.org/10.1103/PhysRevC.93.054911} {\bibfield  {journal}
  {\bibinfo  {journal} {Phys. Rev. C}\ }\textbf {\bibinfo {volume} {93}},\
  \bibinfo {pages} {054911} (\bibinfo {year} {2016})}\BibitemShut {NoStop}%
\bibitem [{\citenamefont {Lin}\ and\ \citenamefont {Zheng}(2021)}]{AMPT2021}%
  \BibitemOpen
  \bibfield  {author} {\bibinfo {author} {\bibfnamefont {Z.-W.}\ \bibnamefont
  {Lin}}\ and\ \bibinfo {author} {\bibfnamefont {L.}~\bibnamefont {Zheng}},\
  }\href {https://doi.org/10.1007/s41365-021-00944-5} {\bibfield  {journal}
  {\bibinfo  {journal} {Nucl. Sci. Tech.}\ }\textbf {\bibinfo {volume} {32}},\
  \bibinfo {pages} {113} (\bibinfo {year} {2021})}\BibitemShut {NoStop}%
\bibitem [{\citenamefont {Wang}\ and\ \citenamefont
  {Gyulassy}(1991)}]{HIJING-1}%
  \BibitemOpen
  \bibfield  {author} {\bibinfo {author} {\bibfnamefont {X.-N.}\ \bibnamefont
  {Wang}}\ and\ \bibinfo {author} {\bibfnamefont {M.}~\bibnamefont
  {Gyulassy}},\ }\href {https://doi.org/10.1103/PhysRevD.44.3501} {\bibfield
  {journal} {\bibinfo  {journal} {Phys. Rev. D}\ }\textbf {\bibinfo {volume}
  {44}},\ \bibinfo {pages} {3501} (\bibinfo {year} {1991})}\BibitemShut
  {NoStop}%
\bibitem [{\citenamefont {Gyulassy}\ and\ \citenamefont
  {Wang}(1994)}]{HIJING-2}%
  \BibitemOpen
  \bibfield  {author} {\bibinfo {author} {\bibfnamefont {M.}~\bibnamefont
  {Gyulassy}}\ and\ \bibinfo {author} {\bibfnamefont {X.-N.}\ \bibnamefont
  {Wang}},\ }\href
  {https://doi.org/https://doi.org/10.1016/0010-4655(94)90057-4} {\bibfield
  {journal} {\bibinfo  {journal} {Comp. Phys. Comm.}\ }\textbf {\bibinfo
  {volume} {83}},\ \bibinfo {pages} {307} (\bibinfo {year} {1994})}\BibitemShut
  {NoStop}%
\bibitem [{\citenamefont {Zhang}(1998)}]{ZPCModel}%
  \BibitemOpen
  \bibfield  {author} {\bibinfo {author} {\bibfnamefont {B.}~\bibnamefont
  {Zhang}},\ }\href
  {https://doi.org/https://doi.org/10.1016/S0010-4655(98)00010-1} {\bibfield
  {journal} {\bibinfo  {journal} {Comp. Phys. Comm.}\ }\textbf {\bibinfo
  {volume} {109}},\ \bibinfo {pages} {193} (\bibinfo {year}
  {1998})}\BibitemShut {NoStop}%
\bibitem [{\citenamefont {Li}\ and\ \citenamefont {Ko}(1995)}]{ARTModel}%
  \BibitemOpen
  \bibfield  {author} {\bibinfo {author} {\bibfnamefont {B.-A.}\ \bibnamefont
  {Li}}\ and\ \bibinfo {author} {\bibfnamefont {C.~M.}\ \bibnamefont {Ko}},\
  }\href {https://doi.org/10.1103/PhysRevC.52.2037} {\bibfield  {journal}
  {\bibinfo  {journal} {Phys. Rev. C}\ }\textbf {\bibinfo {volume} {52}},\
  \bibinfo {pages} {2037} (\bibinfo {year} {1995})}\BibitemShut {NoStop}%
\bibitem [{\citenamefont {Jin}\ \emph {et~al.}(2018)\citenamefont {Jin},
  \citenamefont {Chen}, \citenamefont {Ma} \emph {et~al.}}]{Jin2018}%
  \BibitemOpen
  \bibfield  {author} {\bibinfo {author} {\bibfnamefont {X.-H.}\ \bibnamefont
  {Jin}}, \bibinfo {author} {\bibfnamefont {J.-H.}\ \bibnamefont {Chen}},
  \bibinfo {author} {\bibfnamefont {Y.-G.}\ \bibnamefont {Ma}}, \emph
  {et~al.},\ }\href {https://doi.org/10.1007/s41365-018-0393-1} {\bibfield
  {journal} {\bibinfo  {journal} {Nucl. Sci. Tech.}\ }\textbf {\bibinfo
  {volume} {29}},\ \bibinfo {pages} {54} (\bibinfo {year} {2018})}\BibitemShut
  {NoStop}%
\bibitem [{\citenamefont {Wang}\ \emph {et~al.}(2019)\citenamefont {Wang},
  \citenamefont {Chen}, \citenamefont {Ma} \emph {et~al.}}]{Wang2019}%
  \BibitemOpen
  \bibfield  {author} {\bibinfo {author} {\bibfnamefont {H.}~\bibnamefont
  {Wang}}, \bibinfo {author} {\bibfnamefont {J.~H.}\ \bibnamefont {Chen}},
  \bibinfo {author} {\bibfnamefont {Y.~G.}\ \bibnamefont {Ma}}, \emph
  {et~al.},\ }\href {https://doi.org/10.1007/s41365-019-0706-z} {\bibfield
  {journal} {\bibinfo  {journal} {Nucl. Sci. Tech.}\ }\textbf {\bibinfo
  {volume} {30}},\ \bibinfo {pages} {185} (\bibinfo {year} {2019})}\BibitemShut
  {NoStop}%
\bibitem [{\citenamefont {Wang}\ and\ \citenamefont {Chen}(2022)}]{WangH0}%
  \BibitemOpen
  \bibfield  {author} {\bibinfo {author} {\bibfnamefont {H.}~\bibnamefont
  {Wang}}\ and\ \bibinfo {author} {\bibfnamefont {J.~H.}\ \bibnamefont
  {Chen}},\ }\href {https://doi.org/10.1007/s41365-022-00999-y} {\bibfield
  {journal} {\bibinfo  {journal} {Nucl. Sci. Tech.}\ }\textbf {\bibinfo
  {volume} {33}},\ \bibinfo {pages} {15} (\bibinfo {year} {2022})}\BibitemShut
  {NoStop}%
\bibitem [{\citenamefont {Lin}(2014)}]{AMPT_temperature_parton_Lin}%
  \BibitemOpen
  \bibfield  {author} {\bibinfo {author} {\bibfnamefont {Z.-W.}\ \bibnamefont
  {Lin}},\ }\href {https://doi.org/10.1103/PhysRevC.90.014904} {\bibfield
  {journal} {\bibinfo  {journal} {Phys. Rev. C}\ }\textbf {\bibinfo {volume}
  {90}},\ \bibinfo {pages} {014904} (\bibinfo {year} {2014})}\BibitemShut
  {NoStop}%
\bibitem [{\citenamefont {Shen}\ and\ \citenamefont {Li}(2020)}]{NSTSongFlow}%
  \BibitemOpen
  \bibfield  {author} {\bibinfo {author} {\bibfnamefont {C.}~\bibnamefont
  {Shen}}\ and\ \bibinfo {author} {\bibfnamefont {Y.}~\bibnamefont {Li}},\
  }\href {https://doi.org/https://doi.org/10.1007/s41365-020-00829-z}
  {\bibfield  {journal} {\bibinfo  {journal} {Nucl. Sci. Tech.}\ }\textbf
  {\bibinfo {volume} {31}},\ \bibinfo {pages} {122} (\bibinfo {year}
  {2020})}\BibitemShut {NoStop}%
\bibitem [{\citenamefont {Cao}\ \emph {et~al.}(2022)\citenamefont {Cao},
  \citenamefont {Zhang},\ and\ \citenamefont {Ma}}]{PhysRevC_Cao_2022}%
  \BibitemOpen
  \bibfield  {author} {\bibinfo {author} {\bibfnamefont {R.-X.}\ \bibnamefont
  {Cao}}, \bibinfo {author} {\bibfnamefont {S.}~\bibnamefont {Zhang}},\ and\
  \bibinfo {author} {\bibfnamefont {Y.-G.}\ \bibnamefont {Ma}},\ }\href
  {https://doi.org/10.1103/PhysRevC.106.014910} {\bibfield  {journal} {\bibinfo
   {journal} {Phys. Rev. C}\ }\textbf {\bibinfo {volume} {106}},\ \bibinfo
  {pages} {014910} (\bibinfo {year} {2022})}\BibitemShut {NoStop}%
\bibitem [{\citenamefont {Chen}\ \emph {et~al.}(2023)\citenamefont {Chen},
  \citenamefont {Ma},\ and\ \citenamefont {Chen}}]{Chen_NT}%
  \BibitemOpen
  \bibfield  {author} {\bibinfo {author} {\bibfnamefont {Q.}~\bibnamefont
  {Chen}}, \bibinfo {author} {\bibfnamefont {G.}~\bibnamefont {Ma}},\ and\
  \bibinfo {author} {\bibfnamefont {J.}~\bibnamefont {Chen}},\ }\href
  {https://doi.org/10.11889/j.0253-3219.2023.hjs.46.040013} {\bibfield
  {journal} {\bibinfo  {journal} {Nucl. Tech. (in Chinese)}\ }\textbf {\bibinfo
  {volume} {46}},\ \bibinfo {pages} {040013} (\bibinfo {year}
  {2023})}\BibitemShut {NoStop}%
\bibitem [{\citenamefont {Zhao}\ \emph
  {et~al.}(2019{\natexlab{a}})\citenamefont {Zhao}, \citenamefont {Ma},\ and\
  \citenamefont {Ma}}]{ZhaoXL2019}%
  \BibitemOpen
  \bibfield  {author} {\bibinfo {author} {\bibfnamefont {X.-L.}\ \bibnamefont
  {Zhao}}, \bibinfo {author} {\bibfnamefont {G.-L.}\ \bibnamefont {Ma}},\ and\
  \bibinfo {author} {\bibfnamefont {Y.-G.}\ \bibnamefont {Ma}},\ }\href
  {https://doi.org/10.1103/PhysRevC.99.034903} {\bibfield  {journal} {\bibinfo
  {journal} {Phys. Rev. C}\ }\textbf {\bibinfo {volume} {99}},\ \bibinfo
  {pages} {034903} (\bibinfo {year} {2019}{\natexlab{a}})}\BibitemShut
  {NoStop}%
\bibitem [{\citenamefont {Gao}\ \emph {et~al.}(2020)\citenamefont {Gao},
  \citenamefont {Ma}, \citenamefont {Pu},\ and\ \citenamefont
  {Wang}}]{Gao2020}%
  \BibitemOpen
  \bibfield  {author} {\bibinfo {author} {\bibfnamefont {J.-H.}\ \bibnamefont
  {Gao}}, \bibinfo {author} {\bibfnamefont {G.-L.}\ \bibnamefont {Ma}},
  \bibinfo {author} {\bibfnamefont {S.}~\bibnamefont {Pu}},\ and\ \bibinfo
  {author} {\bibfnamefont {Q.}~\bibnamefont {Wang}},\ }\href
  {https://doi.org/10.1007/s41365-020-00801-x} {\bibfield  {journal} {\bibinfo
  {journal} {Nucl. Sci. Tech.}\ }\textbf {\bibinfo {volume} {31}},\ \bibinfo
  {pages} {90} (\bibinfo {year} {2020})}\BibitemShut {NoStop}%
\bibitem [{\citenamefont {Liu}\ and\ \citenamefont {Huang}(2020)}]{Huang2020}%
  \BibitemOpen
  \bibfield  {author} {\bibinfo {author} {\bibfnamefont {Y.-C.}\ \bibnamefont
  {Liu}}\ and\ \bibinfo {author} {\bibfnamefont {X.-G.}\ \bibnamefont
  {Huang}},\ }\href {https://doi.org/10.1007/s41365-020-00764-z} {\bibfield
  {journal} {\bibinfo  {journal} {Nucl. Sci. Tech.}\ }\textbf {\bibinfo
  {volume} {31}},\ \bibinfo {pages} {56} (\bibinfo {year} {2020})}\BibitemShut
  {NoStop}%
\bibitem [{\citenamefont {Wang}\ \emph {et~al.}(2021)\citenamefont {Wang},
  \citenamefont {Wu}, \citenamefont {Shou}, \citenamefont {Ma}, \citenamefont
  {Ma},\ and\ \citenamefont {Zhang}}]{WangCZ2021}%
  \BibitemOpen
  \bibfield  {author} {\bibinfo {author} {\bibfnamefont {C.-Z.}\ \bibnamefont
  {Wang}}, \bibinfo {author} {\bibfnamefont {W.-Y.}\ \bibnamefont {Wu}},
  \bibinfo {author} {\bibfnamefont {Q.-Y.}\ \bibnamefont {Shou}}, \bibinfo
  {author} {\bibfnamefont {G.-L.}\ \bibnamefont {Ma}}, \bibinfo {author}
  {\bibfnamefont {Y.-G.}\ \bibnamefont {Ma}},\ and\ \bibinfo {author}
  {\bibfnamefont {S.}~\bibnamefont {Zhang}},\ }\href
  {https://doi.org/10.1016/j.physletb.2021.136580} {\bibfield  {journal}
  {\bibinfo  {journal} {Phys. Lett. B}\ }\textbf {\bibinfo {volume} {820}},\
  \bibinfo {pages} {136580} (\bibinfo {year} {2021})}\BibitemShut {NoStop}%
\bibitem [{\citenamefont {Zhao}\ \emph
  {et~al.}(2019{\natexlab{b}})\citenamefont {Zhao}, \citenamefont {Ma},\ and\
  \citenamefont {Ma}}]{Zhao2019}%
  \BibitemOpen
  \bibfield  {author} {\bibinfo {author} {\bibfnamefont {X.~L.}\ \bibnamefont
  {Zhao}}, \bibinfo {author} {\bibfnamefont {G.~L.}\ \bibnamefont {Ma}},\ and\
  \bibinfo {author} {\bibfnamefont {Y.~G.}\ \bibnamefont {Ma}},\ }\href
  {https://doi.org/10.1016/j.physletb.2019.04.002} {\bibfield  {journal}
  {\bibinfo  {journal} {Phys. Lett. B}\ }\textbf {\bibinfo {volume} {792}},\
  \bibinfo {pages} {413} (\bibinfo {year} {2019}{\natexlab{b}})}\BibitemShut
  {NoStop}%
\bibitem [{\citenamefont {Wu}\ \emph {et~al.}(2021)\citenamefont {Wu},
  \citenamefont {Wang}, \citenamefont {Shou}, \citenamefont {Ma},\ and\
  \citenamefont {Zheng}}]{WuWY}%
  \BibitemOpen
  \bibfield  {author} {\bibinfo {author} {\bibfnamefont {W.-Y.}\ \bibnamefont
  {Wu}}, \bibinfo {author} {\bibfnamefont {C.-Z.}\ \bibnamefont {Wang}},
  \bibinfo {author} {\bibfnamefont {Q.-Y.}\ \bibnamefont {Shou}}, \bibinfo
  {author} {\bibfnamefont {Y.-G.}\ \bibnamefont {Ma}},\ and\ \bibinfo {author}
  {\bibfnamefont {L.}~\bibnamefont {Zheng}},\ }\href
  {https://doi.org/10.1103/PhysRevC.103.034906} {\bibfield  {journal} {\bibinfo
   {journal} {Phys. Rev. C}\ }\textbf {\bibinfo {volume} {103}},\ \bibinfo
  {pages} {034906} (\bibinfo {year} {2021})}\BibitemShut {NoStop}%
\bibitem [{\citenamefont {Feldmeier}(1990)}]{Feldmeier:1989st}%
  \BibitemOpen
  \bibfield  {author} {\bibinfo {author} {\bibfnamefont {H.}~\bibnamefont
  {Feldmeier}},\ }\href {https://doi.org/10.1016/0375-9474(90)90328-J}
  {\bibfield  {journal} {\bibinfo  {journal} {Nucl. Phys. A}\ }\textbf
  {\bibinfo {volume} {515}},\ \bibinfo {pages} {147} (\bibinfo {year}
  {1990})}\BibitemShut {NoStop}%
\bibitem [{\citenamefont {Chernykh}\ \emph {et~al.}(2007)\citenamefont
  {Chernykh}, \citenamefont {Feldmeier}, \citenamefont {Neff}, \citenamefont
  {von Neumann-Cosel},\ and\ \citenamefont {Richter}}]{Chernykh:2007zz}%
  \BibitemOpen
  \bibfield  {author} {\bibinfo {author} {\bibfnamefont {M.}~\bibnamefont
  {Chernykh}}, \bibinfo {author} {\bibfnamefont {H.}~\bibnamefont {Feldmeier}},
  \bibinfo {author} {\bibfnamefont {T.}~\bibnamefont {Neff}}, \bibinfo {author}
  {\bibfnamefont {P.}~\bibnamefont {von Neumann-Cosel}},\ and\ \bibinfo
  {author} {\bibfnamefont {A.}~\bibnamefont {Richter}},\ }\href
  {https://doi.org/10.1103/PhysRevLett.98.032501} {\bibfield  {journal}
  {\bibinfo  {journal} {Phys. Rev. Lett.}\ }\textbf {\bibinfo {volume} {98}},\
  \bibinfo {pages} {032501} (\bibinfo {year} {2007})}\BibitemShut {NoStop}%
\bibitem [{\citenamefont {Kanada-En'yo}\ \emph {et~al.}(2012)\citenamefont
  {Kanada-En'yo}, \citenamefont {Kimura},\ and\ \citenamefont
  {Ono}}]{Kanada-Enyo:2012yif}%
  \BibitemOpen
  \bibfield  {author} {\bibinfo {author} {\bibfnamefont {Y.}~\bibnamefont
  {Kanada-En'yo}}, \bibinfo {author} {\bibfnamefont {M.}~\bibnamefont
  {Kimura}},\ and\ \bibinfo {author} {\bibfnamefont {A.}~\bibnamefont {Ono}},\
  }\href {https://doi.org/10.1093/ptep/pts001} {\bibfield  {journal} {\bibinfo
  {journal} {PTEP}\ }\textbf {\bibinfo {volume} {2012}},\ \bibinfo {pages}
  {01A202} (\bibinfo {year} {2012})}\BibitemShut {NoStop}%
\bibitem [{\citenamefont {Kanada-En'Yo}\ \emph {et~al.}(2005)\citenamefont
  {Kanada-En'Yo}, \citenamefont {Kimura},\ and\ \citenamefont
  {Horiuchi}}]{Kanada-Enyo:2005}%
  \BibitemOpen
  \bibfield  {author} {\bibinfo {author} {\bibfnamefont {Y.}~\bibnamefont
  {Kanada-En'Yo}}, \bibinfo {author} {\bibfnamefont {M.}~\bibnamefont
  {Kimura}},\ and\ \bibinfo {author} {\bibfnamefont {H.}~\bibnamefont
  {Horiuchi}},\ }\href {https://doi.org/10.1140/epjad/i2005-06-035-y}
  {\bibfield  {journal} {\bibinfo  {journal} {The European Physical Journal A}\
  }\textbf {\bibinfo {volume} {25}},\ \bibinfo {pages} {305} (\bibinfo {year}
  {2005})}\BibitemShut {NoStop}%
\bibitem [{\citenamefont {He}\ \emph {et~al.}(2016)\citenamefont {He},
  \citenamefont {Ma}, \citenamefont {Cao}, \citenamefont {Cai},\ and\
  \citenamefont {Zhang}}]{He:2016cwt}%
  \BibitemOpen
  \bibfield  {author} {\bibinfo {author} {\bibfnamefont {W.~B.}\ \bibnamefont
  {He}}, \bibinfo {author} {\bibfnamefont {Y.~G.}\ \bibnamefont {Ma}}, \bibinfo
  {author} {\bibfnamefont {X.~G.}\ \bibnamefont {Cao}}, \bibinfo {author}
  {\bibfnamefont {X.~Z.}\ \bibnamefont {Cai}},\ and\ \bibinfo {author}
  {\bibfnamefont {G.~Q.}\ \bibnamefont {Zhang}},\ }\href
  {https://doi.org/10.1103/PhysRevC.94.014301} {\bibfield  {journal} {\bibinfo
  {journal} {Phys. Rev. C}\ }\textbf {\bibinfo {volume} {94}},\ \bibinfo
  {pages} {014301} (\bibinfo {year} {2016})}\BibitemShut {NoStop}%
\bibitem [{\citenamefont {Huang}\ \emph {et~al.}(2017)\citenamefont {Huang},
  \citenamefont {Ma},\ and\ \citenamefont {He}}]{Huang:2017ysr}%
  \BibitemOpen
  \bibfield  {author} {\bibinfo {author} {\bibfnamefont {B.~S.}\ \bibnamefont
  {Huang}}, \bibinfo {author} {\bibfnamefont {Y.~G.}\ \bibnamefont {Ma}},\ and\
  \bibinfo {author} {\bibfnamefont {W.~B.}\ \bibnamefont {He}},\ }\href
  {https://doi.org/10.1103/PhysRevC.95.034606} {\bibfield  {journal} {\bibinfo
  {journal} {Phys. Rev. C}\ }\textbf {\bibinfo {volume} {95}},\ \bibinfo
  {pages} {034606} (\bibinfo {year} {2017})}\BibitemShut {NoStop}%
\bibitem [{\citenamefont {Wang}\ \emph {et~al.}(2015)\citenamefont {Wang},
  \citenamefont {Cao},\ and\ \citenamefont {Zhang}}]{Wangshanshan:2015}%
  \BibitemOpen
  \bibfield  {author} {\bibinfo {author} {\bibfnamefont {S.~S.}\ \bibnamefont
  {Wang}}, \bibinfo {author} {\bibfnamefont {X.~G.}\ \bibnamefont {Cao}},\ and\
  \bibinfo {author} {\bibfnamefont {T.~L.}\ \bibnamefont {Zhang}},\ }\href
  {https://doi.org/10.11804/NuclPhysRev.32.01.024} {\bibfield  {journal}
  {\bibinfo  {journal} {Nucl. Phys. Rev}\ }\textbf {\bibinfo {volume} {32}},\
  \bibinfo {pages} {24} (\bibinfo {year} {2015})}\BibitemShut {NoStop}%
\bibitem [{\citenamefont {Patsyuk}\ \emph {et~al.}(2021)\citenamefont {Patsyuk}
  \emph {et~al.}}]{Patsyuk:2021fju}%
  \BibitemOpen
  \bibfield  {author} {\bibinfo {author} {\bibfnamefont {M.}~\bibnamefont
  {Patsyuk}} \emph {et~al.},\ }\href
  {https://doi.org/10.1038/s41567-021-01193-4} {\bibfield  {journal} {\bibinfo
  {journal} {Nature Phys.}\ }\textbf {\bibinfo {volume} {17}},\ \bibinfo
  {pages} {693} (\bibinfo {year} {2021})}\BibitemShut {NoStop}%
\bibitem [{\citenamefont {Ciofi~degli Atti}\ and\ \citenamefont
  {Simula}(1996)}]{CiofidegliAtti:1995qe}%
  \BibitemOpen
  \bibfield  {author} {\bibinfo {author} {\bibfnamefont {C.}~\bibnamefont
  {Ciofi~degli Atti}}\ and\ \bibinfo {author} {\bibfnamefont {S.}~\bibnamefont
  {Simula}},\ }\href {https://doi.org/10.1103/PhysRevC.53.1689} {\bibfield
  {journal} {\bibinfo  {journal} {Phys. Rev. C}\ }\textbf {\bibinfo {volume}
  {53}},\ \bibinfo {pages} {1689} (\bibinfo {year} {1996})}\BibitemShut
  {NoStop}%
\bibitem [{\citenamefont {Acharya}\ \emph {et~al.}(2018)\citenamefont
  {Acharya}, \citenamefont {Adam} \emph {et~al.}}]{201820}%
  \BibitemOpen
  \bibfield  {author} {\bibinfo {author} {\bibfnamefont {S.}~\bibnamefont
  {Acharya}}, \bibinfo {author} {\bibfnamefont {J.}~\bibnamefont {Adam}}, \emph
  {et~al.},\ }\href
  {https://doi.org/https://doi.org/10.1016/j.physletb.2018.03.051} {\bibfield
  {journal} {\bibinfo  {journal} {Phys. Lett. B}\ }\textbf {\bibinfo {volume}
  {781}},\ \bibinfo {pages} {20} (\bibinfo {year} {2018})}\BibitemShut
  {NoStop}%
\bibitem [{\citenamefont {Li}\ \emph {et~al.}(2022{\natexlab{b}})\citenamefont
  {Li}, \citenamefont {Cruz-Torres}, \citenamefont {Santiesteban} \emph
  {et~al.}}]{eA}%
  \BibitemOpen
  \bibfield  {author} {\bibinfo {author} {\bibfnamefont {S.}~\bibnamefont
  {Li}}, \bibinfo {author} {\bibfnamefont {R.}~\bibnamefont {Cruz-Torres}},
  \bibinfo {author} {\bibfnamefont {N.}~\bibnamefont {Santiesteban}}, \emph
  {et~al.},\ }\href {https://doi.org/10.1038/s41586-022-05007-2} {\bibfield
  {journal} {\bibinfo  {journal} {Nature}\ }\textbf {\bibinfo {volume} {609}},\
  \bibinfo {pages} {41} (\bibinfo {year} {2022}{\natexlab{b}})}\BibitemShut
  {NoStop}%
\bibitem [{\citenamefont {Zhang}\ \emph {et~al.}(2021)\citenamefont {Zhang},
  \citenamefont {Behera}, \citenamefont {Bhatta},\ and\ \citenamefont
  {Jia}}]{Zhang:2021phk}%
  \BibitemOpen
  \bibfield  {author} {\bibinfo {author} {\bibfnamefont {C.}~\bibnamefont
  {Zhang}}, \bibinfo {author} {\bibfnamefont {A.}~\bibnamefont {Behera}},
  \bibinfo {author} {\bibfnamefont {S.}~\bibnamefont {Bhatta}},\ and\ \bibinfo
  {author} {\bibfnamefont {J.}~\bibnamefont {Jia}},\ }\href
  {https://doi.org/10.1016/j.physletb.2021.136702} {\bibfield  {journal}
  {\bibinfo  {journal} {Phys. Lett. B}\ }\textbf {\bibinfo {volume} {822}},\
  \bibinfo {pages} {136702} (\bibinfo {year} {2021})}\BibitemShut {NoStop}%
\bibitem [{\citenamefont {Alver}\ and\ \citenamefont
  {Roland}(2010)}]{Alver:2010gr}%
  \BibitemOpen
  \bibfield  {author} {\bibinfo {author} {\bibfnamefont {B.}~\bibnamefont
  {Alver}}\ and\ \bibinfo {author} {\bibfnamefont {G.}~\bibnamefont {Roland}},\
  }\href {https://doi.org/10.1103/PhysRevC.82.039903} {\bibfield  {journal}
  {\bibinfo  {journal} {Phys. Rev. C}\ }\textbf {\bibinfo {volume} {81}},\
  \bibinfo {pages} {054905} (\bibinfo {year} {2010})}\BibitemShut {NoStop}%
\end{thebibliography}%

\end{document}